\documentclass[12pt]{article}
\usepackage[dvips]{graphicx}
\usepackage{epsfig}
\usepackage[left]{lineno}

\def\NIMA#1#2#3{Nucl. Inst. Methods {\bf A#1} (#2) #3}

\begin{document}

\begin{center}
EUROPEAN ORGANIZATION FOR NUCLEAR RESEARCH\\
\end{center}
\begin{flushright}
CERN-PH-EP/2007-021\\
22 June 2007
\end{flushright}

\begin{center}
\boldmath
{\Large\bf Search for direct CP violating charge
asymmetries in $K^\pm\to\pi^\pm\pi^+\pi^-$ and
$K^\pm\to\pi^\pm\pi^0\pi^0$ decays} \unboldmath
\end{center}

\begin{center}
{\Large The NA48/2 Collaboration}\\
\vspace{2mm}
 J.R.~Batley,
 A.J.~Culling,
 G.~Kalmus,
 C.~Lazzeroni,
 D.J.~Munday,
 M.W.~Slater,
 S.A.~Wotton \\
{\em \small Cavendish Laboratory, University of Cambridge,
Cambridge, CB3 0HE,
U.K.$\,$\footnotemark[1]} \\[0.2cm]
 R.~Arcidiacono,
 G.~Bocquet,
 N.~Cabibbo,
 A.~Ceccucci,
 D.~Cundy$\,$\footnotemark[2],
 V.~Falaleev,
 M.~Fidecaro,
 L.~Gatignon,
 A.~Gonidec,
 W.~Kubischta,
 A.~Norton,
 A.~Maier,
 M.~Patel,
 A.~Peters\\
{\em \small CERN, CH-1211 Gen\`eve 23, Switzerland}
% ...
\\[0.5cm]
\it{Accepted by the European Physics Journal C.} \rm
\end{center}

\setcounter{footnote}{0}
\footnotetext[1]{Funded by the U.K.
Particle Physics and Astronomy Research Council}
\footnotetext[2]{Present address: Istituto di Cosmogeofisica del CNR
di Torino, I-10133 Torino, Italy}
\footnotetext[3]{Present address: Scuola Normale Superiore and INFN,
I-56100 Pisa, Italy}
\footnotetext[4]{Present address: CERN, CH-1211 Gen\`eve 23, Switzerland}
\footnotetext[5]{Also at Centre de Physique des Particules de Marseille,
IN2P3-CNRS, Universit\'e de la M\'editerran\'ee, Marseille, France}
\footnotetext[6] {Also Istituto di Fisica, Universit\`a di Modena,
I-41100 Modena, Italy}
\footnotetext[7]{Istituto di Fisica,
Universit\`a di Urbino, I-61029  Urbino, Italy}
\footnotetext[8]{Present address: Physikalisches Institut,
Universit\"at Bonn, D-53115 Bonn, Germany}
\footnotetext[9]{Funded by the German Federal Minister for Education
and research under contract 05HK1UM1/1}
\footnotetext[10]{Permanent address: SLAC, Stanford University,
Menlo Park, CA 94025, U.S.A.}
\footnotetext[11]{Present address: UCLA, Los Angeles, CA 90024,
U.S.A.}%
\footnotetext[12]{Funded by the German Federal Minister for Research
and Technology (BMBF) under contract 056SI74}
\footnotetext[13]{Funded by the Austrian Ministry for Traffic and
Research under the contract GZ 616.360/2-IV GZ 616.363/2-VIII, and
by the Fonds f\"ur Wissenschaft und Forschung FWF Nr.~P08929-PHY}

%%%%%%%%%%%%%%%%%%%%%%%%%%%%%%%%%%%%%%%%%%%%%%%%%%%%%%
%\begin{linenumbers}

\begin{abstract}
A measurement of the direct CP violating charge asymmetries of the
Dalitz plot linear slopes $A_g=(g^+-g^-)/(g^++g^-)$ in
$K^\pm\to\pi^\pm\pi^+\pi^-$ and $K^\pm\to\pi^\pm\pi^0\pi^0$ decays
by the NA48/2 experiment at CERN SPS is presented. A new technique
of asymmetry measurement involving simultaneous $K^+$ and $K^-$
beams and a large data sample collected allowed a result of an
unprecedented precision. The charge asymmetries were measured to be
$A^c_g=(-1.5\pm2.2)\times10^{-4}$ with $3.11\times 10^9$
$K^{\pm}\to\pi^\pm\pi^+\pi^-$ decays, and
$A^n_g=(1.8\pm1.8)\times10^{-4}$ with $9.13\times 10^7$
$K^{\pm}\to\pi^\pm\pi^0\pi^0$ decays. The precision of the results
is limited mainly by the size of the data sample.
\end{abstract}
\newpage

\section*{Introduction}

For more than 20 years, after the discovery in 1964 that the long
lived neutral kaon could decay to the same $2\pi$ final state as the
short lived one~\cite{ch64}, demonstrating that the mass eigenstates
of the neutral kaons consist of a mixture of even and odd
eigenstates under the combined operation of Charge conjugation and
Parity, no other manifestation of CP violation was detected despite
intensive experimental investigation. In the meanwhile the discovery
of the three families of quarks and the development of the Standard
Model (SM) made it plausible that CP violation was in fact a general
property of the weak interactions, originating from a single
non-trivial phase in the CKM matrix of the coefficients involved in
flavour changing transitions. It was in particular realized that,
barring accidental cancellations, CP violation should be relevant
not only in meson-antimeson mixing but also in decays (direct CP
violation) and in so called mixing induced transitions.

Two major experimental breakthroughs have since taken place. In the
late 1990s, following an earlier indication by NA31~\cite{ba93}, the
NA48 and KTeV experiments firmly established the existence of direct
CP violation~\cite{fa99,al99} by measuring a non-zero value of the
parameter ${\rm Re}(\varepsilon'/\varepsilon)$ parameter in
$K^0\to2\pi$ decays. More recently the B-factory experiments Babar
and Belle discovered a series of CP violating effects in the system
of the neutral $B$ meson~\cite{au01,ab04}.

In order to explore possible non-SM enhancements to heavy-quark
loops, which are at the core of direct CP-violating processes, all
possible manifestations of direct CP violation have to be studied
experimentally. In kaon physics, besides the already investigated
parameter ${\rm Re}(\varepsilon'/\varepsilon)$, the most promising
complementary observables are the rates of GIM-suppressed
flavour-changing neutral current decays $K\to\pi\nu\bar\nu$, and the
charge asymmetry between $K^+$ and $K^-$ decays into $3\pi$.

It is difficult to constrain the fundamental parameters of the
theory using measurements of direct CP violation in decay amplitudes
due to the presence of non-perturbative hadron effects. Still, an
intense theoretical programme is under way to improve predictions,
aiming to allow the direct CP violation measurements to be used as
quantitative constraints on the SM.

The $K^\pm\to3\pi$ matrix element squared is conventionally
parameterized~\cite{pdg} by a polynomial expansion\footnote{At the
next order of approximation, electromagnetic interactions and final
state $\pi\pi$ rescattering should be included into the decay
amplitude~\cite{ca04,ca05,co06}. These effects, despite being
charge-symmetric, may still contribute to the interpretation of the
results, as will be discussed below.}
\begin{equation}
|M(u,v)|^2\sim 1+gu+hu^2+kv^2, \label{slopes}
\end{equation}
where $g$, $h$, $k$ are the so called linear and quadratic Dalitz
plot slope parameters ($|h|,|k|\ll |g|$), and the two Lorentz
invariant kinematic variables $u$ and $v$ are defined as
\begin{equation}
u=\frac{s_3-s_0}{m_\pi^2},~~v=\frac{s_2-s_1}{m_\pi^2},~~
s_i=(P_K-P_i)^2,~i=1,2,3;~~s_0=\frac{s_1+s_2+s_3}{3}. \label{uvdef}
\end{equation}
Here $m_\pi$ is the charged pion mass, $P_K$ and $P_i$ are the kaon
and pion four-momenta, the indices $i=1,2$ correspond to the two
pions of the same electrical charge (``even'' pions, so that $v$ is
defined up to a sign), and the index $i=3$ to the pion of different
charge (the ``odd'' pion). A difference between the slope parameters
$g^+$ and $g^-$ describing the decays of positive and negative
kaons, respectively, is a manifestation of direct CP violation
usually expressed by the corresponding slope asymmetry
\begin{equation}
A_g = (g^+ - g^-)/(g^+ + g^-) \approx \Delta g/(2g), \label{agdef}
\end{equation}
where $\Delta g$ is the slope difference and $g$ is the average
linear slope. In general terms, the slope asymmetry is expected to
be strongly enhanced with respect to the asymmetry of integrated
decay rates~\cite{is92}. A recent full next-to-leading order ChPT
computation~\cite{ma95} predicts $A_g$ to be of the order of
$10^{-5}$ within the SM. Another SM calculation~\cite{fa05} predicts
the asymmetry in the $K^\pm\to\pi^\pm\pi^0\pi^0$ decay to be of the
order of $10^{-6}$ (in agreement with~\cite{ma95} within the
errors). Theoretical calculations involving processes beyond the
SM~\cite{sh98} allow a wider range of $A_g$, including substantial
enhancements up to a few $10^{-4}$.

In the past years, several experiments have searched for the CP
violating slope asymmetry in both $\pi^\pm\pi^+\pi^-$ and
$\pi^\pm\pi^0\pi^0$ decay modes by collecting samples of $K^+$ and
$K^-$ decays~\cite{fo70,sm75}. These measurements set upper limits
on $A_g$ at the level of a few $10^{-3}$, limited by systematic
uncertainties.

The primary goal of the NA48/2 experiment at the CERN SPS is the
measurement of the slope charge asymmetries $A_g$ in both
$K^\pm\to\pi^\pm\pi^+\pi^-$ and $K^\pm\to\pi^\pm\pi^0\pi^0$
processes with a sensitivity at least one order of magnitude better
than previous experiments. The new level of precision can explore
effects, albeit larger than the SM predictions, induced by new
physics, and is achieved by using a novel measurement technique
based on simultaneous $K^+$ and $K^-$ beams overlapping in space.

Measurements of $A_g$ in both decay modes, performed with
approximately half of the NA48/2 data sample have been
published~\cite{k3pi,k3pi-n}. This paper presents the final results,
superseding these earlier results.

The plan of the paper is the following. In Section 1 the
experimental setup is described. Section 2 contains the description
of the method developed for the measurement of the slope difference
which is common to both decay modes, and which aims to cancel first
order systematic biases. The analyses of the
$K^\pm\to\pi^\pm\pi^+\pi^-$ and $K^\pm\to\pi^\pm\pi^0\pi^0$ decay
modes are discussed in Sections 3 and 4, respectively. Discussion of
the results follows in Section 5.

%%%%%%%%%%%%%%%%%%%%%%%%%%%%%%%%%%%%%%%%%%%%%%%%%%%%%%%
\section{Beams and detectors}

\boldmath
\subsection{Simultaneous $K^+$ and $K^-$ beams}
\unboldmath

\begin{figure}[tb]
\vspace{-6mm}
\begin{center}
{\resizebox*{\textwidth}{!}{\includegraphics{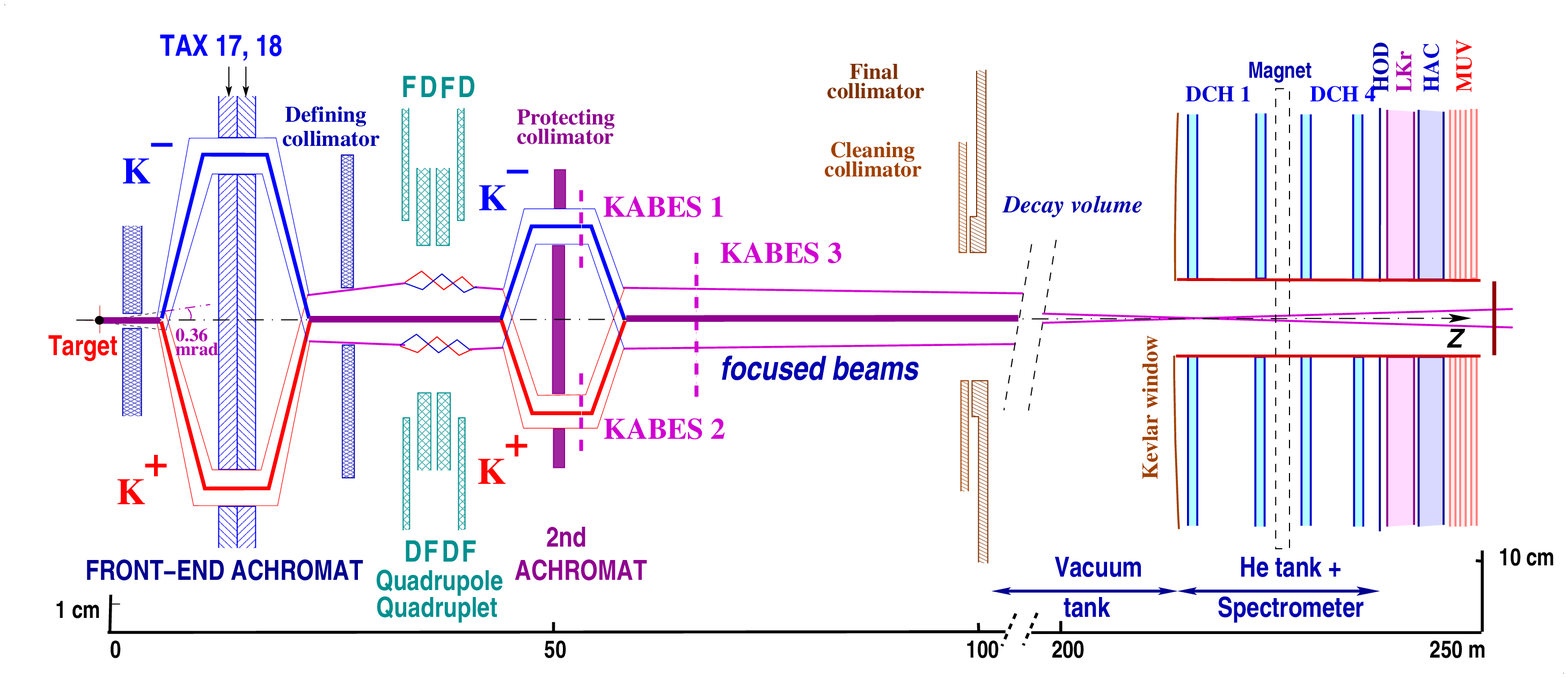}}}
\end{center}
\vspace{-6mm} \caption{Schematic side view of the NA48/2 beam line
(TAX17,18: motorized beam dump/collimators used to select the
momentum of the $K^+$ and $K^-$ beams, FDFD/DFDF: focusing
quadrupoles, KABES1--3: kaon beam spectrometer stations), decay
volume and detector (DCH1--4: drift chambers, HOD: hodoscope, LKr:
EM calorimeter, HAC: hadron calorimeter, MUV: muon veto). Thick
lines indicate beam axes, narrow ones the projection of their
envelopes. Note that the vertical scales are different in the two
parts of the figure.} \label{fig:beams}
\end{figure}

A high precision measurement of $A_g$ (at the level of $10^{-4}$)
requires a dedicated experimental approach together with collection
of very large data samples. A novel beam line providing for the
first time two simultaneous charged beams of opposite signs
superimposed in space over the decay fiducial volume was designed
and built in the high intensity hall (ECN3) at the CERN SPS. The
beam line is a key element of the experiment, as it allows decays of
$K^+$ and $K^-$ to be recorded at the same time, and therefore leads
to cancellation of several systematic uncertainties for the charge
asymmetry measurement. Regular alternation of magnetic fields in all
the beam line elements was adopted, which contributed to
symmetrization of the average geometrical acceptance for $K^+$ and
$K^-$ decays. The layout of the beams and detectors is shown
schematically in Fig.~\ref{fig:beams}.

The setup is described in a right-handed orthogonal coordinate
system with the $z$ axis directed downstream along the beam, and the
$y$ axis directed vertically up. Thus the horizontal $x$ coordinate
is such that the negative (positive) $x$ point to the Sal\`eve
(Jura) mountains. The Sal\`eve/Jura notation is used in the
following analysis to denote the spectrometer magnet polarity.

The beams are produced by 400 GeV/$c$ protons (delivered from the
CERN SPS with a duty cycle of 4.8~s/16.8~s) impinging on a beryllium
target of 40~cm length and 2~mm diameter. Both beams leave the
target on axis at zero production angle, thereby ensuring that their
source is geometrically similar and that the $K^+/K^-$ flux ratio
remains stable. It was demonstrated that the small residual
difference of $K^+$ and $K^-$ momentum spectra produces a negligible
effect on the measurement. Charged particles with momenta $(60\pm3)$
GeV/$c$ are selected in a charge-symmetric way by an achromatic
system of four dipole magnets with zero total deflection
(`achromat'), which splits the two beams in the vertical plane and
then recombines them on a common axis. Then the beams pass through a
defining collimator and a series of four quadrupoles designed to
produce horizontal and vertical charge-symmetric focusing of the
beams towards the detector. Finally the two beams are again split in
the vertical plane and recombined by a second achromat, where three
stations of a Micromegas-type~\cite{gi96} detector operating in TPC
mode form a kaon beam spectrometer KABES~\cite{pe04} (not used in
the present analyses).

After passing through the cleaning and final collimators, the beams
enter the decay volume housed in a 114 m long cylindrical vacuum
tank with a diameter of 1.92 m for the first 65 m, and 2.4 m for the
rest. Both beams follow the same path in the decay volume: their
axes coincide to 1~mm, while their lateral sizes are about 1~cm.
With $7\times 10^{11}$ protons per burst incident on the target, the
positive (negative) beam flux at the entrance of the decay volume is
$3.8\times 10^7$ ($2.5\times 10^7$) particles per pulse (primarily
charged pions), of which $5.7\%$ ($4.9\%$) are $K^+$ ($K^-$). The
$K^+/K^-$ flux ratio is about $1.8$ and stable to within 1\%; no
correlations were observed between its variation in time and
time-dependent inefficiencies of the setup. The fraction of beam
kaons decaying in the decay volume at nominal momentum is $22\%$.

\subsection{Main detectors}

The decay volume is followed by a magnetic spectrometer~\cite{bed95}
used to reconstruct tracks of charged particles. The spectrometer is
housed in a tank filled with helium at nearly atmospheric pressure,
separated from the vacuum tank by a thin ($0.31\%X_0$) Kevlar
composite window. A thin-walled aluminium beam pipe of 16 cm outer
diameter traversing the centre of the spectrometer and all the
following detectors allows the undecayed beam particles and the muon
halo from decays of beam pions to continue their path in vacuum. The
spectrometer consists of four identical drift chambers (DCH): DCH1,
DCH2 located upstream, and DCH3, DCH4 downstream of a dipole magnet.
The magnet has a field integral $\int B_ydz=0.4$~Tm, providing a
horizontal transverse momentum kick $\Delta P_x=120~{\rm MeV}/c$ for
charged particles. The DCHs have the shape of a regular octagon with
transverse size of about 2.8 m and fiducial area of about 4.5 m$^2$.
Each chamber is composed of eight planes of sense wires arranged in
four pairs of staggered planes oriented horizontally, vertically,
and along each of the two orthogonal $45^\circ$ directions. The
spatial resolution of each DCH is $\sigma_x=\sigma_y=90~\mu$m. The
nominal momentum resolution of the magnetic spectrometer is
$\sigma_p/p = (1.02 \oplus 0.044\cdot p)\%$ ($p$ expressed in
GeV/$c$). The measured resolution on the reconstructed $3\pi^\pm$
invariant mass varied during the data taking time in the range of
$(1.65-1.72)$~MeV/$c^2$ in 2003 data, and $(1.76-1.82)$~MeV/$c^2$ in
2004, depending on DCH performance (generally, the chambers were
operated at lower high voltage in 2004).

The magnetic spectrometer is followed by a plastic scintillator
hodoscope (HOD) used to produce fast trigger signals and to provide
precise time measurements of charged particles. The HOD, with a
transverse size of about 2.4 m, consists of a plane of vertical and
a plane of horizontal strip-shaped counters, each plane comprising
64 counters arranged in four quadrants. Each quadrant is logically
subdivided into 4 subquadrants (``segments'') which take part in the
trigger logic. Counter widths (lengths) vary from 6.5 cm (121 cm)
for central counters to 9.9 cm (60 cm) for peripheral ones.

The HOD is followed by a liquid krypton electromagnetic calorimeter
(LKr)~\cite{ba96} used to detect electrons and photons. It is an
almost homogeneous ionization chamber with an active volume of
7~m$^3$ of liquid krypton, segmented transversally into 13248
projective cells, 2$\times$2 cm$^2$ each, by a system of Cu$-$Be
ribbon electrodes, and with no longitudinal segmentation. The
calorimeter is $27X_0$ deep and has an energy resolution
$\sigma(E)/E=0.032/\sqrt{E}\oplus0.09/E\oplus0.0042$ ($E$ is
expressed in GeV). The spatial resolution for a single
electromagnetic shower is
$\sigma_x=\sigma_y=0.42/\sqrt{E}\oplus0.06$ cm for the transverse
coordinates $x$ and $y$.

The LKr is followed by a hadronic calorimeter (HAC) and a muon
detector (MUV), both not used in the present analyses. A detailed
description of the components of the NA48 detector can be found
elsewhere~\cite{fa07}.

\newpage
\subsection{Trigger logic}
\label{sec:triggers}

The event rate of $\sim 500$~kHz is dominated by
$K^\pm\to\mu^\pm\nu$ and $K^\pm\to\pi^\pm\pi^0$ decays, which are of
limited physics interest as such within the NA48/2 programme. A
two-level trigger system is used to select the $K^\pm\to3\pi$ decay
modes for readout, reducing the event rate to $\sim 10$~kHz.

At the first level (L1), the $K^\pm\to\pi^\pm\pi^+\pi^-$ decays are
triggered by requiring coincidences of hits in the two HOD planes in
at least two of the 16 segments (the L1C condition). The
$K^\pm\to\pi^\pm\pi^0\pi^0$ decays are triggered by requiring a
coincidence of the two HOD planes in at least one segment, and the
presence of at least two distinguishable clusters of energy
deposition in the LKr (the L1N condition). In some periods of data
taking the L1N condition also used as input the total LKr energy
deposition (see Section~\ref{sec:neutrig}). The L1C signal is
produced by the HOD logic, while the L1N signal consists of HOD and
LKr components. These two components require separate analyses in
order to study possible biases to the asymmetry measurement.

The second level trigger (L2) is based on a real time system
computing coordinates of DCH hits using DCH drift times, and a farm
of asynchronous microprocessors performing a fast reconstruction of
tracks and running a decision-taking algorithm.

The L2 algorithm selecting the $K^\pm\to\pi^\pm\pi^0\pi^0$ events
(L2N) examines the events passing the L1N condition, and requires
the existence of a reconstructed track which, assumed to be a pion,
has an energy $E^*<230$ MeV in the rest frame of a $K^\pm$ having a
momentum of 60 GeV/$c$ directed along the $z$ axis. This condition
suppresses the $K^\pm\to\pi^\pm\pi^0$ decays (which have $E^*=248$
MeV), while keeping the $K^\pm\to\pi^\pm\pi^0\pi^0$ decays (for
which $E^*$ ranges between 140 MeV and 193 MeV).

The L2 algorithm selecting the $K^\pm\to\pi^\pm\pi^+\pi^-$ events
(L2C) examines the events passing the L1C condition, and requires at
least two tracks to originate in the decay volume with the
reconstructed distance of closest approach below 5 cm. L1C triggers
not satisfying this condition are examined further and accepted if
the L2N condition is satisfied.

NA48/2 collected data during two runs in 2003 and 2004, with
$\sim$50 days of efficient data taking in each run. About $18\times
10^9$ triggers, and 200 TB of data were recorded in total.

%%%%%%%%%%%%%%%%%%%%%%%%%%%%%%%%%%%%%%%%%%%%%%%%%%%%%%%%%%%%%%
\section{The method of the slope difference measurement}

\subsection{Data taking strategy}
\label{sec:strategy}

Charge symmetrization of the experimental conditions is to a large
extent achieved by using simultaneous superimposed $K^+$ and $K^-$
beams with similar momentum spectra. However, the presence of
magnetic fields in both the beam line (achromats, focusing
quadrupoles, etc.) and the magnetic spectrometer, combined with some
asymmetries in detector performance, introduces residual charge
asymmetries. In order to equalize the local effects on the
acceptance, the polarities of all the magnets in the beam transport
system were reversed during the data taking on an approximately
weekly basis (corresponding to the periodicity of SPS technical
stops), while the polarity of the spectrometer magnet was alternated
on a more frequent basis (approximately once per day in 2003 and
once every 3 hours in 2004).

Data collected over a period which has all four possible setup
configurations (i.e. combinations of beam line and spectrometer
magnet polarities), spanning about two weeks of efficient data
taking, represent a ``supersample'' and is treated as an independent
and self-consistent set of data for the asymmetry measurement.

For the $K^\pm\to\pi^\pm\pi^+\pi^-$ analysis, nine supersamples
numbered 0 to 8 were collected in two years of data taking
(supersamples 0--3 in 2003 and supersamples 4--8 in 2004). For the
$K^\pm\to\pi^\pm\pi^0\pi^0$ analysis, a fraction of the data sample
was rejected due to poor trigger performance, and another fraction
of data was merged into a larger supersample to improve the balance
of magnet polarities, which resulted in seven supersamples numbered
I to VII (supersamples I--III in 2003 and supersamples IV--VII in
2004).

\subsection{Fitting procedure}
\label{sec:fitpro}

The $u$-projection of the Dalitz plot is sufficient to extract the
information about $\Delta g$ at the desired level of precision. The
measurement method is based on comparing the reconstructed
$u$-spectra of $K^+$ and $K^-$ decays (denoted as $N^+(u)$ and
$N^-(u)$, respectively). In the framework of the parameterization
(\ref{slopes}), the ratio $R(u)=N^+(u)/N^-(u)$ is in good
approximation proportional to
\begin{equation}
R(u) \sim 1+\frac{\Delta g~\!u}{1+gu+hu^2},
\end{equation}
the contribution of the $kv^2$ term integrated over $v$ being
negligible. The slope difference $\Delta g$ can be extracted from a
fit to $R(u)$ involving the measured slope parameters $g$ and
$h$~\cite{pdg,slopes}, and $A_g$ can be evaluated as in
Eq.~(\ref{agdef}).

The parameters describing the Dalitz plot distribution explicitly
appear in the fitting function. This generates a certain dependence
of the results on the assumed shape of the event density, which in
our case is Eq.~(\ref{slopes}). These effects will be discussed
later.

\subsection{Cancellation of systematic effects}

It should be noted that any instrumental effect can induce a fake
slope difference only if it is: (1) charge asymmetric and (2)
correlated with $u$.

For a given decay mode, each supersample contains four sets of
simultaneously collected $K^+\to3\pi$ and $K^-\to3\pi$ samples
corresponding to the four different setup configurations (eight data
samples in total). To measure the charge asymmetry, exploiting the
cancellations of systematic biases emerging due to polarity
reversals, the following ``quadruple ratio'' $R_4(u)$ is evaluated.
It involves the eight corresponding $u$ spectra, and is formed by
the product of four $R(u)=N^+(u)/N^-(u)$ ratios with opposite kaon
sign and a deliberately chosen setup configuration in numerator and
denominator:
\begin{equation}
R_4(u) = R_{US}(u)\cdot R_{UJ}(u)\cdot R_{DS}(u)\cdot R_{DJ}(u).
\label{quad}
\end{equation}
Here the indices $U$/$D$ denote beam line polarities corresponding
to $K^+$ passing along the upper/lower path in the achromats of the
beamline, while the indices $S$/$J$ denote spectrometer magnet
polarities (opposite for $K^+$ and $K^-$) corresponding to the
``even'' (i.e. the two identical) pions from a
$K^\pm\to\pi^\pm\pi^+\pi^-$ decay being deflected to
negative/positive $x$ (i.e. towards the Sal\`eve/Jura mountains).
For example, $R_{US}(u)$ is the ratio of the $u$ distribution for
$K^+$ transported along the upper path in the beamline achromats and
collected with a certain polarity of the spectrometer magnetic
field, to the distribution for $K^-$ transported along the lower
path and collected with the opposite analyzing magnet polarity. A
fit of the quadruple ratio~(\ref{quad}) with a function of the form
\begin{equation}
f(u)=n\cdot\left(1+\frac{\Delta g~\!u}{1+gu+hu^2}\right)^4
\label{flin4}
\end{equation}
results in the determination of two parameters: the normalization
$n$ and the difference of slopes $\Delta g$. The normalization is
sensitive to the $K^+/K^-$ flux ratio, while $\Delta g$ is not.

The rationale for choosing the four ratios $R(u)$ appearing
in~(\ref{quad}) as the basic ones (which is not the only
possibility) lies in the fact that they intrinsically cancel at
first order instrumental effects linked to the imperfect left-right
symmetry of the apparatus. As will be seen below, time variation of
the left-right asymmetry (which is primarily due to variations of
spectrometer misalignment and beam geometry) is the largest
instrumental effect, and has been the primary subject of the
analysis.

The quadruple ratio technique logically completes the procedure of
magnet polarity reversal, and allows a three-fold cancellation of
systematic biases in the data, without the need to rely on an
accurate simulation of the instrumental asymmetries:
\begin{itemize}
\item due to spectrometer magnet polarity reversal, local detector
  inefficiencies cancel between $K^+$ and $K^-$ samples with decay products
  reaching the same parts of the detector in each of the four ratios
  $R(u)$ appearing in the quadruple ratio $R_4(u)$;
\item due to the simultaneous beams, global time-variable biases cancel
  between $K^+$ and $K^-$ samples recorded at the same time
  in the product of $R_S(u)$ and $R_J(u)$ ratios;
\item due to beam line polarity reversal, local beam line biases,
  resulting in slight differences in beam profiles and momentum spectra,
  largely cancel between the $R_U(u)$ and $R_D(u)$ ratios.
\end{itemize}

The method is independent of the $K^+/K^-$ flux ratio and the
relative sizes of the samples collected with different setup
configurations. However, the statistical precision is limited mainly
by the smallest of the samples involved, therefore the balance of
sample sizes was controlled during the data taking. The result
remains sensitive only to time variations of asymmetries in the
experimental conditions which have a characteristic time smaller
than the corresponding field alternation period.

\subsection{Control quantities}

In order to demonstrate that the level of cancellation of the
systematic uncertainties achieved using the quadruple ratio
technique is sufficient, the quantities cancelling in (\ref{quad})
have to be measured. For this purpose, slopes of two other control
quadruple ratios built out of the eight $u$ spectra are evaluated.
These control ratios can be written as the products of the four
ratios of the $u$ spectra for same sign kaons recorded with
different setup configurations. As a result, any physical asymmetry
cancels in these ratios, while the setup asymmetries do not.

The fake slope difference $\Delta g_{SJ}$ introduced by global
time-dependent detector variations does not cancel in a ratio with
opposite spectrometer polarities and identical beam line polarities
in numerator and denominator, or equivalently, in the adopted
notation
\begin{equation}
\label{quad_LR}
R_{SJ}(u) = (R_{US}(u)\cdot R_{DS}(u)) /
(R_{UJ}(u)\cdot R_{DJ}(u)). \label{lr}
\end{equation}
Similarly, the fake slope difference $\Delta g_{UD}$ introduced by
the differences of the two beam paths does not cancel in a ratio
with opposite beam line polarities and identical spectrometer
polarities, namely:
\begin{equation}
\label{quad_UD}
R_{UD}(u) = (R_{US}(u)\cdot R_{UJ}(u)) / (R_{DS}(u)\cdot R_{DJ}(u)).
\label{ud}
\end{equation}
It should be noted that time stability of the beam line conditions
is much better than the (itself small) difference between the upper
and lower beam paths.

The intrinsic left-right asymmetry of the experimental apparatus
which, as already mentioned, cancels in each of the basic ratios
appearing in~(\ref{quad}), can not be directly measured by the above
two control ratios (although it is accessible within an analysis
based in different choice of the basic ratios). These asymmetric
effects, being the largest of the residual effects and the central
subject of investigation, have been taken into account with
dedicated methods (see Sections~\ref{sec:alignment},
\ref{sec:beam-geom}).

Any possible systematic bias remaining in (\ref{quad}) is of a
higher order effect than $\Delta g_{SJ}$ and $\Delta g_{UD}$. Thus,
a measurement of fake slope differences compatible with zero, within
their statistical uncertainties, validates the measurement method.

\subsection{Averaging over the independent data sets}
\label{sec:averaging}

As described in Section~\ref{sec:strategy}, the data sample is
divided into several independent self-contained supersamples. Two
methods of averaging the independent results over the supersamples
were considered:
\begin{itemize}
\item averaging the measured quadruple ratios (\ref{quad}) independently
for each $u$ bin, and then fitting the resulting grand quadruple
ratio with the function (\ref{flin4});
\item independent fitting of the quadruple ratios in every
supersample, and then averaging the results on $\Delta g$.
\end{itemize}
The first method allows the results to be presented in a compact
form, i.e. in terms of a single spectrum for each of the two decay
modes. On the other hand, the second method allows a cross-check
concerning several aspects of the analysis, in particular the time
stability of the results.

As will be demonstrated below, the two methods lead to similar
results due to the good balance of statistics which was maintained
between the supersamples.

\subsection{Monte Carlo simulation}

Due to the method described above, no Monte Carlo (MC) corrections
to the acceptance are required. Nevertheless, a detailed
GEANT-based~\cite{geant} MC simulation was developed as a tool for
the studies of possible systematic effects; this includes full
detector geometry and material description, stray magnetic fields
simulation, local DCH inefficiencies, DCH misalignment, and the beam
lines simulation (which allows for a reproduction of kaon momentum
spectra and beam profiles). Moreover, time variations of the above
effects during the running period were simulated.

A large-scale MC production was carried out for both $K^\pm\to3\pi$
decay modes, providing samples of sizes comparable to those of the
data. Namely, $\sim 10^{10}$ events were generated for each decay
mode with the correct balance for each beam and detector
configuration closely matching that of the data.

%%%%%%%%%%%%%%%%%%%%%%%%%%%%%%%%%%%%%%%%%%%%%%%%%%%%%%%%%%%
%\newpage
\boldmath
\section{Slope difference in $K^\pm\to3\pi^\pm$ decay}
\unboldmath

\subsection{Event reconstruction and selection}

Reconstruction of $K^\pm\to3\pi^\pm$ events is based on the magnetic
spectrometer information. Tracks are reconstructed from hits in DCHs
using the measured magnetic field map rescaled to the recorded value
of the electric current in the spectrometer analyzing magnet.
Systematic uncertainties arising from this procedure due to
spectrometer misalignment and imperfect knowledge of magnetic fields
are discussed in Sections~\ref{sec:alignment}
and~\ref{sec:spectrometer_field}, respectively.

Three-track vertices, compatible with the $K^\pm\to3\pi^\pm$ decay,
are reconstructed using the Kalman filter algorithm~\cite{fr87} by
extrapolation of track segments from the upstream part of the
spectrometer into the decay volume, taking into account multiple
scattering in the Kevlar window and helium volume, and the stray
magnetic field in the decay volume due to the Earth's field and
parasitic magnetization of the vacuum tank.

The stray field is non-uniform, and has a typical magnitude of 0.5~G
(comparable to the Earth field); the field map was measured in the
entire vacuum tank before the 2003 run. Its effect is to induce a
transverse deviation of about 1~mm and an angular deviation of about
$10^{-5}$ rad to a 20 GeV/$c$ charged particle traversing
longitudinally the whole decay volume. Accounting for the stray
field in the vertex reconstruction reduces the amplitude of the
measured sinusoidal variation of the reconstructed $3\pi^\pm$
invariant mass with respect to the azimuthal orientation of the odd
pion by more than an order of magnitude, to a level below 0.05
MeV/$c^2$. The event kinematics is calculated using the
reconstructed track momenta and directions as extrapolated to the
decay vertex.

Several stages of compaction and filtering of the data were applied,
reducing the data volume from $200$ TB to $1.23$ TB, while reducing
the number of events in the sample from $18\times10^9$ to
$4.87\times10^9$. The applied filtering algorithm includes rejection
of low quality data and also soft kinematic constraints: the main
requirement is the presence of at least 3 reconstructed tracks in
the event, which rejects about $55\%$ of the recorded triggers.

The principal selection criteria applied to the reconstructed
variables are listed below. Note that corrections to track momenta
for spectrometer misalignment and magnetic field, discussed in the
following Sections~\ref{sec:alignment}
and~\ref{sec:spectrometer_field}, are applied before the kinematics
of the event is reconstructed and the selection is made.
\begin{itemize}
\item Total charge of the three pion candidates: $Q=\pm1$.
\item Total transverse momentum with respect to the $z$ axis:
$P_T<0.3$~GeV/$c$, consistent with the angular spread of the beams
and spectrometer resolution.
\item Longitudinal vertex position $Z_{vtx}$ is within the decay volume:
$Z_{vtx}>Z_{fc}$, where $Z_{fc}$ is the longitudinal coordinate of
the final collimator of the beamline. This condition is imposed
since the stray magnetic fields upstream of the final collimator
have not been measured, and therefore appropriate corrections can
not be made. An upper cut on $Z_{vtx}$ is not imposed, as the
geometric acceptance diminishes to zero towards the spectrometer by
itself due to the presence of the beam pipe.
\item Transverse decay vertex position within the beam spot:
its distance from the $z$ axis $R_{vtx}<3$~cm.
\item Consistent track timing from DCHs: $|t_i-t_{avg}|<10$ ns
for each track $i=1,2,3$, where $t_{avg}=(t_1+t_2+t_3)/3$, which
leads to a reasonably low event pile-up probability.
\item Reconstructed kaon momentum is required to be
consistent with the beam momentum spectrum: $54~{\rm GeV}/c<|\vec
P_K|<66~{\rm GeV}/c$.
\item Reconstructed $3\pi$ invariant mass:
$|M_{3\pi}-M_K|<9$~MeV/$c^2$, where $M_K$ is the PDG charged kaon
mass~\cite{pdg}.
\end{itemize}

\begin{figure}[tb]
\begin{center}
\resizebox{0.49\textwidth}{!}{\includegraphics{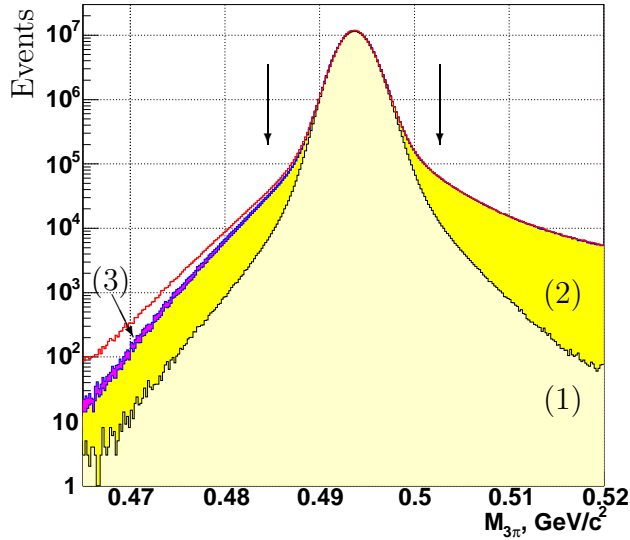}}
\put(-33,50){(1)}\put(-33,90){(2)}\put(-204,95){(3)}
\put(-196,93){\vector(1,-2){8}}
\put(-137,180){\vector(0,-1){30}}\put(-72,180){\vector(0,-1){30}}
\put(-234,167){\rotatebox{90}{Events}}
\end{center}
\vspace{-8mm} \caption{Reconstructed spectrum of $3\pi^\pm$
invariant mass (upper envelope curve) and its comparison to
normalized MC components: (1) events without $\pi\to\mu\nu$ decay in
flight, (2) events with $\pi\to\mu\nu$ decay, (3) radiative
$K_{3\pi\gamma}$ events. The selection conditions are marked with
vertical arrows. The deficit of MC events in the low mass area is
mostly outside the signal region.} \label{fig:kmass}
\end{figure}

The distribution of the $3\pi^\pm$ invariant mass (before the cut on
that quantity) and its comparison to MC are presented in
Fig.~\ref{fig:kmass}. The non-Gaussian tails of the mass
distribution are mainly due to $\pi^\pm\to\mu^\pm\nu_\mu$ decay in
flight (the spectrometer reconstructing the resulting
muon)\footnote{The shape and size of the non-Gaussian tails notably
depend on the adopted cuts on vertex transverse position and total
transverse momentum.}, which is charge-symmetric. The tails are well
understood in terms of MC simulation, and are considered as part of
the signal. A deficit of MC events observed in the low mass region
does not influence the analysis, since it is mostly outside the
signal region; moreover the analysis does not rely on the MC for
acceptance computation. Due to the absence of background and the
presence of a non-Gaussian contribution of pion decay the selection
condition for $M_{3\pi}$ corresponds to five times the resolution.

This principal stage of selection leaves a sample of $3.82\times
10^9$ $K^\pm\to3\pi^\pm$ events which is practically background
free, as $K^\pm\to3\pi^\pm$ is the dominant decay mode of the
charged kaon with more than one charged particle in the final state.
The fact that backgrounds due to other decays of beam kaons and
pions are negligible was confirmed by a MC simulation.\vspace{2mm}

\subsection{Correction for spectrometer misalignment}
\label{sec:alignment}

The transverse positions of DCHs and individual wires were measured
and realigned at the level of reconstruction software every 2--4
weeks of data taking using data collected during special alignment
runs in which muon tracks were recorded with no magnetic field in
the spectrometer. This allows an alignment precision of
$\sim30~\mu$m to be reached. Spectrometer misalignment itself can
not bias the asymmetry measurement. However, time variations of DCH
alignment on a short time scale potentially can, since an
uncorrected shift of a DCH along the $x$ axis leads to
charge-antisymmetric mismeasurement of the momenta.

An unambiguous measure\footnote{Assuming CPT conservation.} of the
residual transverse horizontal misalignment is the difference
between the average reconstructed $3\pi$ invariant masses
corresponding to decays of $K^+$ and $K^-$, denoted as
$\Delta\overline M$. The following sensitivities were determined
from the data: a shift of the DCH4 along the $x$ axis by 1 $\mu$m
with respect to its nominal position induces a measured mass
difference of $\Delta\overline M=1.4$~keV/$c^2$; a difference of
1~$\mu$m between the DCH4 $x$-positions between the data sets taken
with opposite spectrometer polarities induces a fake slope
difference of $\delta(\Delta g^c)=0.03\times 10^{-4}$. Sensitivities
to shifts of the other DCHs are similar.

Monitoring of $\Delta\overline M$ revealed significant transverse
movements of the DCHs between the individual alignment runs, at a
rate of typically below $\sim5\mu$m/day and never exceeding
20~$\mu$m/day\footnote{On an exceptional occasion during the 2003
data taking, a DCH position shifted by 200$\mu$m in one day.}, which
introduces spurious slope differences of the order of a few units of
$10^{-5}$. Introduction of time-dependent corrections to the
measured momenta based on the observed $\Delta\overline M$ reduces
these fake slope differences by more than an order of magnitude to a
negligible level of $\delta(\Delta g^c)<0.1\times 10^{-4}$.

\subsection{Effects due to spectrometer magnetic field}
\label{sec:spectrometer_field}

The measurement of pion momenta is based on the knowledge of the
magnetic field in the spectrometer magnet. The variation of the
current in the magnet biases the overall momentum scale of the
spectrometer. This variation can be directly measured with a
relative precision of $5\times 10^{-4}$; smaller variations are
continuously monitored with a precision of $\sim 10^{-5}$ using the
deviation of the reconstructed charge-averaged kaon mass from the
nominal PDG value~\cite{pdg}. A time-dependent correction can be
introduced by scaling the reconstructed track momenta symmetrically
for positive and negative tracks. However, the momentum scale
effects are a priori highly charge-symmetric by design, due to the
simultaneous $K^+$ and $K^-$ beams (this was also explicitly
verified by comparing the results obtained with and without the
correction). Therefore no correction was applied for spectrometer
current.

On the contrary, the effects caused by a non-uniform permanent (not
inverting with the spectrometer magnetic field polarity) component
of the magnetic field in the region of the spectrometer magnet are
potentially charge asymmetric. They were studied by artificially
introducing the corresponding distortions to measured track momenta
depending on the coordinates of impact points in the magnetic plane,
consistent with the measurement precision of the magnetic field map
and the expected size of the permanent field ($\sim 1$~G). The
resulting variation of the result of $\delta(\Delta g^c)=0.3\times
10^{-4}$ was considered as a residual systematic uncertainty due to
this effect.

Remarkably, a statistically significant $\Delta g^c$ measured with
the events from the side bands of $3\pi^\pm$ mass distribution (i.e.
outside the signal region) can be achieved by introducing certain
realistic configurations of a non-uniform permanent magnetic field
in the region of the spectrometer magnet.

\subsection{Correction for instability of beam geometry}
\label{sec:beam-geom}

The geometric acceptance for the $K^\pm\to\pi^\pm\pi^+\pi^-$ decays
is mainly determined by the vacuum beam pipe traversing the centres
of the DCHs, and the material in the central region of each DCH
where certain groups of DCH wires terminate\footnote{Due to a
relatively small $Q$-value of the $K^\pm\to3\pi^\pm$ decay, $Q=75.0$
MeV, the outer edges of the DCHs do not bias the acceptance.}.
Moreover, the beam optics can only control the average transverse
beam positions to $\pm1$~mm. Time variations of the transverse beam
positions within the mentioned precision generate a sizable
charge-asymmetric bias to the acceptance inducing instrumental slope
asymmetries of the order of a few units of $10^{-4}$. It would
require a stability of the transverse beam positions to the level of
$100~\mu$m in order to reduce the bias to a negligible level.
However it is possible to determine the average $K^+$ and $K^-$ beam
positions as functions of time to this order of precision and apply
charge symmetric cuts, as explained below.

Inner DCH geometrical acceptance cuts which fully contain the beam
pipe and the surrounding DCH regions are applied to the positions of
pion impact points $\vec R_{\pi i}^{1,4}$ in the planes of DCH1 and
DCH4 relative to the average beam intercepts in the DCH planes.
These vary slightly with time and differ for $K^+$ and $K^-$.

The transverse coordinates of a beam kaon $\vec R_{0}^{1,2}$ in the
planes of DCH1 and DCH2 for each event are reconstructed as the
momentum-weighted averages of the coordinates $\vec R_{\pi i}^{1,2}$
of the three reconstructed pions: $\vec R_{0}^{1,2}=\sum_{i=1}^3
(\vec R_{\pi i}^{1,2}|\vec P_{\pi i}|)/\sum_{i=1}^3 |\vec P_{\pi
i}|$, where $|\vec P_{\pi i}|$ is reconstructed momentum of a pion.
Transverse coordinates of a beam kaon in the plane of DCH4 $\vec
R_{0}^{4}$ corresponding to absence of the bending by the analyzing
magnet are computed by linear extrapolation using $\vec R_{0}^{1}$
and $\vec R_{0}^{2}$.

The average beam positions in the planes of DCH1 and DCH4
$\langle\vec R_{0}^{1,4}\rangle$ are computed on the basis of the
distributions of $\vec R_{0}^{1}$ and $\vec R_{0}^{4}$ for the
selected event sample. A bias introduced to $\langle\vec
R_{0}^{1,4}\rangle$ by the fact that $|\vec R_{\pi i}^{1,2}|$ and
$|\vec P_{\pi i}|$ are themselves affected by the acceptance is
negligible. A database of the average beam positions $\langle\vec
R_{0}^{1,4}\rangle$ depending on kaon sign, time (excursions of
$\sim 1$ mm), kaon momentum (excursions of $\sim 1$ mm in the
horizontal plane, $\sim 1$ cm in the vertical plane), and time
within SPS spill (excursions of $\sim 1$ mm) was created.

The conditions $|\vec R_{\pi i}^{1,4}-\langle\vec
R_{0}^{1,4}\rangle|>11.5$~cm, $i=1,2,3$, are applied to symmetrize
the beam geometry effects. These cuts cost $12\%$ of the statistics,
leading to a sample of $3.36\times 10^9$ events. The minimum
distance of 11.5 cm is chosen to ensure that the region of the beam
pipe and the adjacent central insensitive areas of the DCHs are
securely excluded by the cut.

The residual systematic effects arise from the stray magnetic field
in the decay volume, which deflects $\vec R_{\pi i}^{1,4}$ with
respect to $\langle\vec R_{0}^{1,4}\rangle$ in a
charge-antisymmetric way. Corrections for the stray field to the
measured $\vec R_{0}^{1,4}$ were performed. However, the precision
of these corrections is limited by the precision of magnetic field
measurement, which leads to a residual systematic uncertainty of
$\delta(\Delta g^c)=0.2\times 10^{-4}$.

\subsection{Correction for trigger inefficiency}
\label{sec:trigcor}

Only charge-asymmetric trigger inefficiencies correlated with $u$
can possibly bias the measurement. Inefficiencies of the individual
trigger components were directly measured as functions of $u$ using
control data samples from prescaled low bias triggers collected
along with the main ones. This allowed for an accounting for time
variations of the efficiencies, and for a propagation of the
statistical errors of the measured inefficiencies into the final
result.

The trigger logic is described in Section~\ref{sec:triggers}. The
control trigger condition for the L1 efficiency measurement requires
at least one coincidence of hits in the two planes of the HOD. The
control triggers for the L2 efficiency measurement are L1 triggers
recorded regardless of the L2 response. The statistics of each of
the two control samples is roughly 1\% of the main
sample\footnote{Sizes of the control samples are adequate to measure
trigger inefficiencies with a precision better that the statistical
error, due to sufficiently low trigger inefficiencies.}.

The L1 trigger condition requires the coincidence of hits in two of
the 16 non-overlapping HOD segments. This condition is loose as
there are three charged particles in a fully reconstructed event,
and the resulting inefficiency is low. It was measured to be
$0.9\times 10^{-3}$ and found to be stable in time. Due to a few
short-term malfunctions of a HOD channel, several subsamples of the
data sample are affected by higher inefficiency (up to $7\times
10^{-3}$), the source of the inefficiency being localized in space.
This kind of inefficiency was reduced (and symmetrized) in the
selected data sample by applying appropriate geometric cuts to the
pion impact points on the hodoscope surface for the relevant
supersamples. This procedure led to the loss of 7.1\% of the
statistics, reducing the final sample to $3.11\times 10^9$ events.
Due to good time stability of the L1 inefficiency, no bias from the
L1 trigger is assumed. An overall uncertainty of the L1 bias was
conservatively estimated to $\delta(\Delta g^c)=0.3\times 10^{-4}$,
limited by the statistics of the control sample.

Two components of the L2 trigger inefficiency were identified: one
due to trigger timing misalignment, and the other due to local DCH
inefficiency (so called ``geometrical''). The part related to timing
misalignment has a size of $\sim 0.2\%$, and a priori does not
affect the result, being uncorrelated to the kinematic variables.
Its charge symmetry was checked with a detailed MC simulation of
pile-up effects, and a study of the dependence of the result on the
number of allowed accidental tracks. On the contrary, the
geometrical part is correlated to event kinematics, and varies in
time due to variations of the local DCH inefficiencies. These
inefficiencies affect the trigger more than the offline
reconstruction due to lower redundancy and worse online resolutions.
For this reason the measured $u$ spectra are corrected for this part
of the inefficiency. The size of the inefficiency for the selected
sample was measured to be close to $0.6\times 10^{-3}$, but some
periods are affected by higher inefficiency of up to 1.5\% (its
sources not being localized in space in a simple way). The
correction to the whole statistics amounts to $\Delta(\Delta
g^c)=(-0.1\pm0.3)\times 10^{-4}$, where the error is statistical
owing to the limited size of the control sample.

The above procedure leads to over-estimation of the systematic
uncertainties related to trigger inefficiencies, since the
correlations of trigger inefficiencies in bins of $u$ (in other
words, smoothness of variation of trigger efficiency over $u$) are
not taken into account.

\boldmath
\subsection{Fits to $\Delta g^c$ and cross checks}
\unboldmath

\begin{figure}[tb]
\begin{center}
\resizebox{0.99\textwidth}{!}{\includegraphics{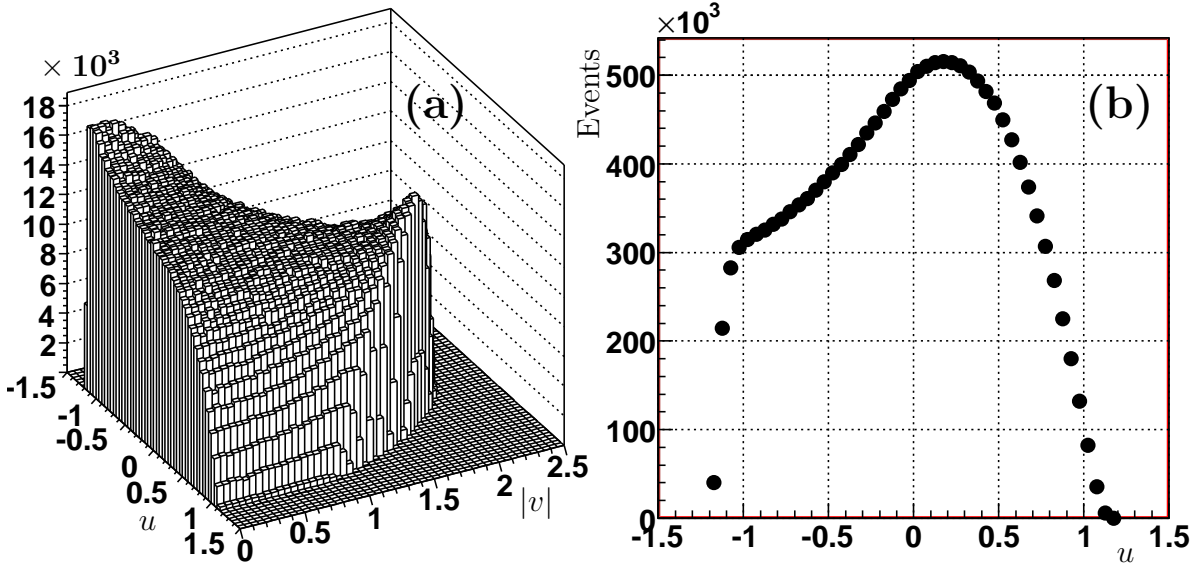}}
\put(-300,178){\Large\bf (a)} \put(-42,178){\Large\bf (b)}
\put(-256,29){$|v|$}\put(-400,22){$u$} \put(-30,10){$u$}
\put(-440,194){$\bf\times 10^3$}
\put(-234,170){\rotatebox{90}{Events}}
\end{center}
\vspace{-8mm} \caption{(a) Reconstructed distribution of the
selected $K^\pm\to\pi^\pm\pi^+\pi^-$ events in the kinematic
variables ($u$,$|v|$); (b) its projection to the $u$ axis. The
distributions correspond to a fraction of the data sample.}
\label{fig:uv}
\end{figure}

The reconstructed Dalitz plot distribution of the events passing the
selection and the corrections described in the previous Sections
(corresponding to a fraction of the sample) is presented in
Fig.~\ref{fig:uv}a. Its projection to the $u$ axis is presented in
Fig.~\ref{fig:uv}b\footnote{The eight $u$ spectra corresponding to
various combinations of kaon sign and magnetic field polarities
involved into the computation are not presented separately, since
the corresponding systematic biases are small, and the differences
between the spectra are difficult to see by eye.}.

The quadruple ratios of the $u$ spectra (\ref{quad}) were computed,
and $\Delta g^c$ was measured by the two methods described in
Section~\ref{sec:averaging} by fitting with the function
(\ref{flin4}). The values of slope parameters $g^c=-0.21134$,
$h^c=0.01848$ recently measured by NA48/2~\cite{slopes} using 55\%
of the 2003 data sample, and consistent with the world averages,
were used. The uncertainties on the values of the above slope
parameters lead to negligible effects on the result.

\begin{figure}[tb]
\begin{center}
{\resizebox*{0.7\textwidth}{!}{\includegraphics{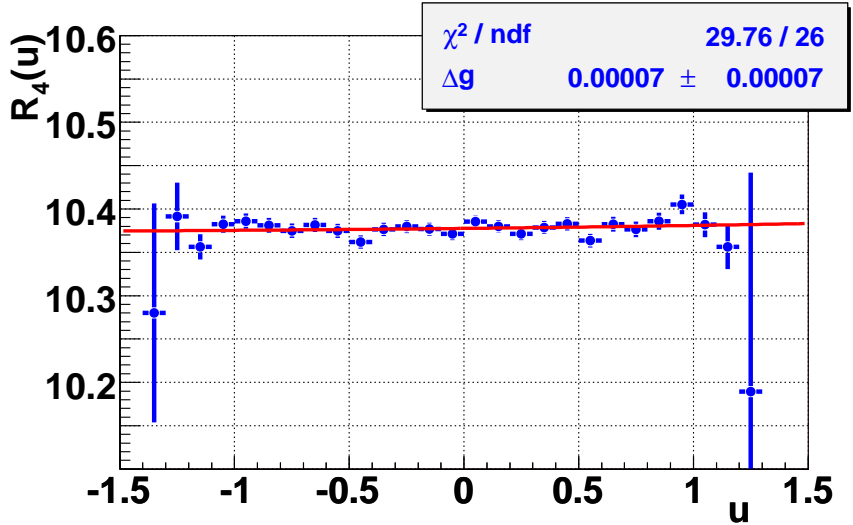}}}
\end{center}
\vspace{-9mm} \caption{The quadruple ratio $R^c_4(u)$ corrected for
L2 trigger efficiency averaged in bins of $u$ over supersamples
fitted with the function (\ref{flin4}) with $g^c=-0.21134$,
$h^c=0.01848$~\cite{slopes}. The point in the first bin, given in
the Table~\ref{tab:fit}, is out of the vertical range.}
\label{fig:fit}
\end{figure}

The grand quadruple ratio obtained by averaging quadruple ratios
over supersamples in each bin of $u$ (corrected for L2 trigger
efficiency) is tabulated in Table~\ref{tab:fit}, and presented along
with the result of the corresponding fit in Fig.~\ref{fig:fit}.

\begin{table}
\caption{The quadruple ratio $R^c_4(u)$ corrected for L2
inefficiency averaged over the supersamples.}
\begin{center}
\begin{tabular}{rrr|rrr}
\hline
$u$ bin centre & Content & Error &$u$ bin centre & Content & Error\\
\hline
$-1.45$~~~~ &  8.96034 & 0.52320 &$-0.05$~~~~ & 10.37140 & 0.00730\\
$-1.35$~~~~ & 10.28018 & 0.12639 & $0.05$~~~~ & 10.38555 & 0.00729\\
$-1.25$~~~~ & 10.39128 & 0.03900 & $0.15$~~~~ & 10.37973 & 0.00730\\
$-1.15$~~~~ & 10.35639 & 0.01459 & $0.25$~~~~ & 10.37140 & 0.00741\\
$-1.05$~~~~ & 10.38233 & 0.00983 & $0.35$~~~~ & 10.37868 & 0.00755\\
$-0.95$~~~~ & 10.38593 & 0.00883 & $0.45$~~~~ & 10.38286 & 0.00777\\
$-0.85$~~~~ & 10.38105 & 0.00855 & $0.55$~~~~ & 10.36338 & 0.00805\\
$-0.75$~~~~ & 10.37476 & 0.00834 & $0.65$~~~~ & 10.38241 & 0.00849\\
$-0.65$~~~~ & 10.38148 & 0.00815 & $0.75$~~~~ & 10.37654 & 0.00907\\
$-0.55$~~~~ & 10.37467 & 0.00798 & $0.85$~~~~ & 10.38572 & 0.00992\\
$-0.45$~~~~ & 10.36193 & 0.00780 & $0.95$~~~~ & 10.40517 & 0.01137\\
$-0.35$~~~~ & 10.37647 & 0.00764 & $1.05$~~~~ & 10.38179 & 0.01434\\
$-0.25$~~~~ & 10.37972 & 0.00747 & $1.15$~~~~ & 10.35615 & 0.02568\\
$-0.15$~~~~ & 10.37670 & 0.00734 & $1.25$~~~~ & 10.18942 & 0.25256\\
\hline
\end{tabular}
\end{center}
\label{tab:fit}
\end{table}

\begin{figure}[p]
\begin{center}
\resizebox{0.81\textwidth}{!}{\includegraphics{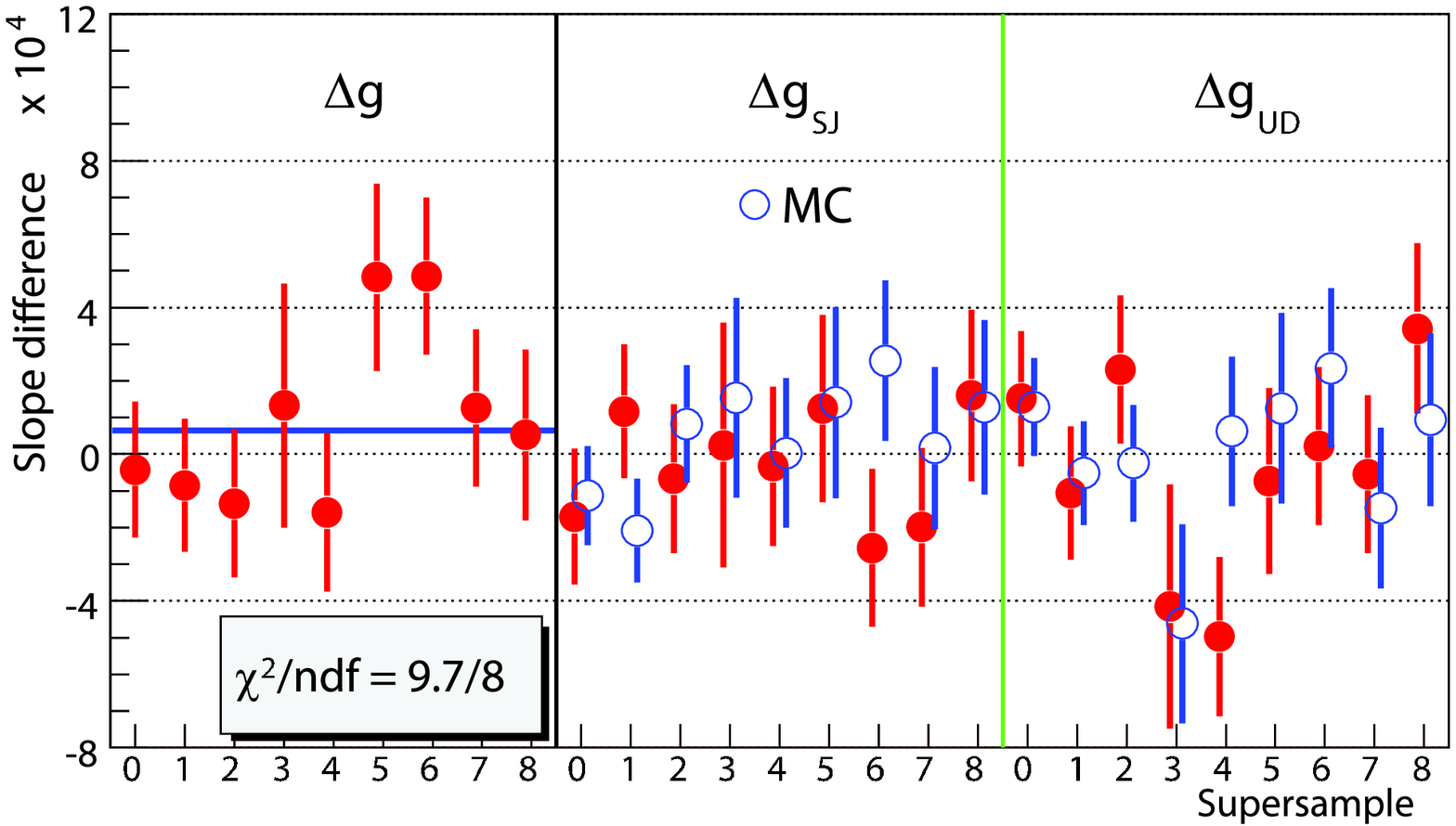}}
\put(-336,31){\Large\bf (a)} \put(-222,31){\Large\bf (b)}
\put(-110,31){\Large\bf (c)}
\end{center}
\vspace{-10mm} \caption{(a) $\Delta g^c$ measurement in the nine
supersamples; control
  quantities (b) $\Delta g_{SJ}^c$ and (c) $\Delta g_{UD}^c$ corresponding
  to detector and beam line asymmetries which cancel in the quadruple
  ratio, and their comparison to MC (open circles).}
\label{fig:stabplot}
\end{figure}

The results of the independent fits for each supersample, including
the numbers of events selected, the ``raw'' values of $\Delta g^c$
obtained without applying the trigger corrections, and the final
values of $\Delta g^c$ with the L2 trigger corrections applied are
presented in Table~\ref{tab:stats}. The independent results obtained
for the nine supersamples are shown in Fig.~\ref{fig:stabplot}(a):
the individual measurements of $\Delta g^c$ are statistically
compatible with a $\chi^2/{\rm ndf} = 9.7/8$.

\begin{table}[p]
\caption{Statistics selected in each supersample and measured
$\Delta g^c$: ``raw'' and corrected for L2 trigger inefficiency. The
errors are statistical only; the errors in the last column include
the L2 trigger efficiency errors.}
\begin{center}
\begin{tabular}{r|r|r|r|r}
\hline Supersample & $K^+\!\to\pi^+\pi^+\pi^-\!\!$ &
$K^-\!\to\pi^-\pi^-\pi^+\!\!$
&$\Delta g^c\times 10^4$ & $\Delta g^c\times 10^4$\\
&decays in $10^6\!\!$&decays in $10^6\!\!$&raw~~~~~&corrected\\
\hline
0 & 448.0 & 249.7 & $ 0.7\pm1.4$ & $-0.4\pm1.8$\\
1 & 270.8 & 150.7 & $-0.8\pm1.8$ & $-0.8\pm1.8$\\
2 & 265.5 & 147.8 & $-1.4\pm2.0$ & $-1.3\pm2.0$\\
3 &  86.1 &  48.0 & $ 0.6\pm3.2$ & $ 1.3\pm3.3$\\
4 & 232.5 & 129.6 & $-2.7\pm1.9$ & $-1.6\pm2.2$\\
5 & 142.4 &  79.4 & $ 5.0\pm2.5$ & $ 4.8\pm2.6$\\
6 & 193.8 & 108.0 & $ 4.9\pm2.1$ & $ 4.9\pm2.2$\\
7 & 195.9 & 109.1 & $ 1.4\pm2.1$ & $ 1.3\pm2.1$\\
8 & 163.9 &  91.4 & $ 1.4\pm2.3$ & $ 0.5\pm2.3$\\
\hline
Total & 1998.9 & 1113.7 & $0.8\pm0.7$ & $0.7\pm0.7$\\
\hline
\end{tabular}
\end{center}
\label{tab:stats}
\end{table}

The measured control quantities $\Delta g_{SJ}^c$ and $\Delta
g_{UD}^c$, which are the slopes of the control ratios
(\ref{quad_LR}) and (\ref{quad_UD}), for the nine supersamples are
presented in Fig.~\ref{fig:stabplot}(b) and \ref{fig:stabplot}(c),
respectively (data points are overlayed with MC ones). The sizes of
these slopes induced by residual time-variable imperfections of the
apparatus which cancel in the result (\ref{quad}) are of the same
order of magnitude as the statistical errors ($10^{-4}$), indicating
that second order effects which could induce non-zero values for
them are negligible. Moreover, the comparison with MC simulations
shows that the sizes of the apparatus asymmetries are well
understood in terms of local inefficiencies and variations of beam
optics.

\subsection{Residual systematic effects}
\label{sec:resid-charged}

Effects due to the difference of cross sections of $\pi^+$ and
$\pi^-$ hadronic interactions with the material of the detector were
evaluated by a simulation of $K\to3\pi$ decays taking into account a
parameterized energy dependence of cross sections of $\pi^\pm N$
interactions~\cite{pdg,cu02} and the material composition of the
detector. As the most striking example of the effect of the charge
asymmetry of $\pi^\pm N$ cross sections, a $\pi^-p$ interaction in
the first plastic scintillator plane of the HOD, giving rise to a
fully neutral final state, produces a trigger bias for
$K^-\to\pi^-\pi^0\pi^0$ decays relative to $K^+\to\pi^+\pi^0\pi^-$
decays. Owing to the kaon momentum spectrum, $\pi^\pm$ momentum in
the detector is restricted to $p_\pi>5$~GeV/$c$, which 1) validates
the use of the cross section parameterization, and 2) makes the
measurement insensitive to the largest differences between $\pi^+p$
and $\pi^-p$ cross sections occurring at $p_{\pi}\sim 1$~GeV/$c$.
The integral charge-asymmetric effects were found to be of the order
of $10^{-4}$, however their $u$-dependence was found to be
negligible, inducing a bias of only the order of $\Delta g\sim
10^{-6}$.

The effects due to the difference of $K^+$ and $K^-$ production
spectra by the primary protons~\cite{at80} do not cancel in the
quadruple ratio. The difference between the $K^+$ and $K^-$ spectra
was quantified by measuring the slope of the quadruple ratio of the
reconstructed $K^+$ and $K^-$ momentum spectra. The slope of the
$K^+$/$K^-$ spectra ratio $f(p)$ normalized by $f(60~{\mathrm
GeV})=1$ was determined to be $df(p)/dp=0.6\%/$(GeV/$c$). This
induces charge asymmetry of geometrical acceptance leading to a fake
slope difference of $\delta(\Delta g^c)=0.3\times 10^{-3}$, as
estimated by a MC simulation. This value was conservatively taken as
the corresponding systematic uncertainty.

Taking into account that the composition of the beams is not charge
symmetric (in particular, the $K^+$ and $K^-$ fluxes differ), event
distortions caused by pile-up with the products of another kaon
decay or a beam halo particle traversing the sensitive region of the
spectrometer is a potential source of systematic bias. To study the
pile-up effects an accidental activity generator was introduced into
the MC. This generator was tuned using the measured composition of
beam and halo fluxes, and a production of $\sim10^8$ correlated
pairs of an original kaon decay event and a piled-up event was
carried out. No charge-asymmetric effects were observed in the
reconstructed $u$ distributions nor in the L2 trigger inefficiencies
down to a level of $\delta(\Delta g^c)=0.2\times 10^{-4}$, limited
by MC statistics.

Biases due to resolution effects were studied by using various
methods of expressing the $u$ variable in terms of directly measured
quantities (using only the invariant mass of the pair of even pions
in the laboratory frame; using the energy of the odd pion in the
kaon rest frame; using a 3C kinematic fit constraining kaon mass and
direction) differing in resolution as a function of $u$. Stability
of the result with respect to variation of bin size in $u$ has been
studied as well. An estimate for the systematic uncertainty due to
resolution effects of $\delta(\Delta g^c)=0.2\times 10^{-4}$ has
been obtained.

%\newpage
\boldmath
\subsection{The resulting $A_g^c$}
\unboldmath

A summary of the systematic uncertainties, including a correction
for L2 trigger inefficiency, is presented in Table~\ref{tab:syst}.
The difference in the linear slope parameter of the Dalitz plot of
the $K^\pm\to\pi^\pm\pi^+\pi^-$ decays, measured with the full
NA48/2 data sample of $3.11\times 10^9$ events, is found to be
\begin{equation}
\Delta g^c = g^+-g^- = (0.7 \pm 0.7_{stat.} \pm 0.4_{trig.} \pm
0.6_{syst.})\times 10^{-4}.
\end{equation}
Here the individual systematic errors are added in quadrature, and
the errors due to trigger inefficiencies are of statistical nature.
Converted to the direct CP violating charge asymmetry (\ref{agdef})
using the value of the Dalitz plot slope $g^c=-0.21134\pm0.00017$
recently measured by the NA48/2~\cite{slopes},
\begin{equation}
A_g^c = (-1.5 \pm 1.5_{stat.} \pm 0.9_{trig.} \pm 1.3_{syst.})\times
10^{-4} = (-1.5\pm2.2)\times 10^{-4}.
\end{equation}

\begin{table}[tb]
\caption{Systematic uncertainties and
  the correction for L2 trigger inefficiency for $\Delta g^c$ measurement.}
\begin{center}
\begin{tabular}{l|c}
\hline
Systematic effect & Correction, uncertainty $\delta(\Delta g^c)\times 10^4$\\
\hline
Spectrometer misalignment               & $\pm0.1$\\
Spectrometer magnetic field             & $\pm0.3$\\
Beam geometry and stray magnetic fields & $\pm0.2$\\
Kaon production spectra                 & $\pm0.3$\\
Pile-up                                 & $\pm0.2$\\
Resolution and fitting                  & $\pm0.2$\\
\hline
Total purely systematic uncertainty     & $\pm0.6$\\
\hline
L1 trigger inefficiency            & $\pm0.3$\\
L2 trigger inefficiency            & $-0.1\pm0.3$\\
\hline
\end{tabular}
\end{center}
\label{tab:syst}
\end{table}

%%%%%%%%%%%%%%%%%%%%%%%%%%%%%%%%%%%%%%%%%%%%%%%%%%%%%%%%%%%
\boldmath
\section{Slope difference in $K^\pm\to\pi^\pm\pi^0\pi^0$ decay}
\unboldmath

\subsection{Event reconstruction and selection}

The $K^\pm\to\pi^\pm\pi^0\pi^0$ decays are reconstructed considering
$\pi^0\to\gamma\gamma$ decays
%\footnote{The decay
%$\pi^0\to\gamma\gamma$ is the dominant decay mode of a $\pi^0$ with
%$BR=98.8\%$; the smallness of $\pi^0$ mean life
%($\tau=8.4\times10^{-17}$~s) makes the pion decay vertex
%experimentally indistinguishable from the kaon decay vertex.}
of each of the $\pi^0$s and hence the reconstruction of the four
photons is required. The principal selection criteria are described
below.
\begin{itemize}
\item An energy deposition cluster in the LKr
is considered to correspond to a photon candidate if the following
conditions are fulfilled: 1) it has an energy $E>3$~GeV, which
minimizes effects of nonlinearity of the LKr response (typically
2$\%$ at 3 GeV and becoming negligible above 10 GeV); 2) it is
situated at distances larger than 10 cm from other clusters, and at
distances larger than 15 cm from impact points of the reconstructed
charged particles, which minimizes effects of energy sharing between
the reconstructed clusters due to overlap; 3) it satisfies
requirements on distances from the outer LKr edges and the beam
pipe, which ensures full lateral containment of the electromagnetic
showers.
\item The event is required to have at least one reconstructed track of a
charged particle, and at least four photon candidates.
\item To suppress the charge-asymmetric
DCH acceptance bias induced by the time instability of beam
geometry, cuts are applied to distances between the track impact
points in DCH1 and DCH4 planes $\vec R_{\pi}^{1,4}$ and the average
reconstructed beam positions $\langle\vec R_{0}^{1,4}\rangle$. These
cuts are similar to those applied in the $K^\pm\to\pi^\pm\pi^+\pi^-$
analysis; the rationale and a description are contained in
Section~\ref{sec:beam-geom}.
\end{itemize}

For each selected event, a $K^\pm\to\pi^\pm\pi^0\pi^0$ decay is
reconstructed as follows. Assuming that a pair of photon candidates
$i,j$ ($i,j=$1,2,3,4) originates from a $\pi^0\to\gamma\gamma$ decay
occurring at a distance $D_{ij}$ from the LKr front face, then
$D_{ij}$ is calculated to very good approximation as
$D_{ij}=R_{ij}\sqrt{E_iE_j}/m_{\pi^0}$, where $E_i$ and $E_j$ are
the energies of the $i$-th and $j$-th photon candidates, $R_{ij}$ is
the distance between their impact points at the LKr front plane, and
$m_{\pi^0}$ is the PDG $\pi^0$ mass~\cite{pdg}.

To search for two $\pi^0\to\gamma\gamma$ decays occurring at the
same point of the decay volume, among all the combinations of
non-overlapping photon candidate pairs ($i,j$) and ($k,l$) the one
with the smallest value of $|D_{ij}-D_{kl}|$ is selected. Moreover,
the smallest of $|D_{ij}-D_{kl}|$ is required to be less than
500~cm, while the resolution on the difference $D_{ij}-D_{kl}$ for
photon pairs originating from the same point of space is $\sim
100$~cm. For the best selected (if any) combination ($i,j$) and
($k,l$), the value of $(D_{ij}+D_{kl})/2$ is used to define the
longitudinal position of a $K^\pm$ decay vertex $Z_{vtx}$.

No geometrical information about the $\pi^\pm$ track is used for
vertex reconstruction in order to avoid the related
charge-asymmetric biases induced by beam geometry variation and
stray magnetic fields.

The following selection criteria are applied to the reconstructed
event kinematics.
\begin{itemize}
\item The longitudinal
vertex position required to be within the decay volume:
$Z_{vtx}>Z_{fc}$, where $Z_{fc}$ is the longitudinal coordinate of
the final collimator.
\item Consistent photon and track timing: $|t^\gamma_{avg}
-t^\gamma_i|<5$, $|t^\gamma_{avg} -t^\pm|<20$ ns, where $t^\gamma_i$
are times of the four selected LKr clusters, $t^\pm$ is the time of
the selected track, and $t^\gamma_{avg}=\sum t^\gamma_i/4$.
\item Reconstructed kaon momentum is required to be consistent with
the beam momentum spectrum: $54~{\rm GeV}/c<|\vec P_K|<66~{\rm
GeV}/c$.
\item Reconstructed $3\pi$ invariant mass:
$|M_{3\pi}-M_K|<6$~MeV/$c^2$. This cut is narrower than in the
$K^\pm\to3\pi^\pm$ case due to a better mass resolution.
\end{itemize}

\begin{figure}[tb]
\begin{center}
\vspace{-9mm}
\resizebox{.5\textwidth}{!}{\includegraphics{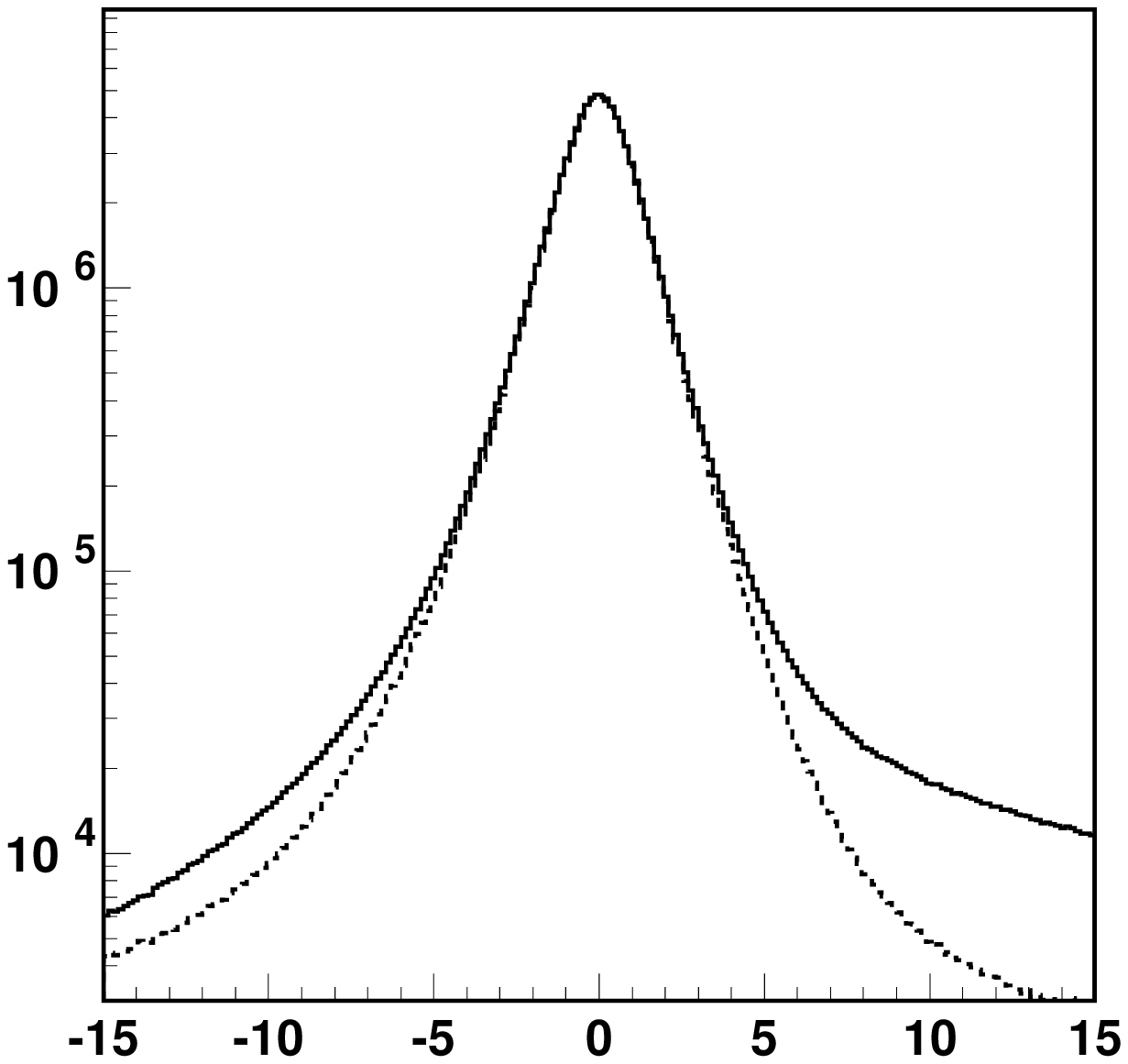}}
\put(-150,9){$m_{3\pi}-m_K^{PDG}$, MeV/$c^2$}
\put(-225,160){\rotatebox{90}{Events}}
\end{center}
\vspace{-10mm} \caption{Deviation of the reconstructed
$\pi^\pm\pi^0\pi^0$ invariant mass from the PDG kaon
mass~\cite{pdg}. The dashed histogram shows the same distribution
for events with no hits in the muon detector (not used in the
analysis).} \label{neutral-mass}
\end{figure}

The above requirements lead to the final sample of $9.13\times 10^7$
events. Fig.~\ref{neutral-mass} shows the $\pi^\pm\pi^0\pi^0$
invariant mass distribution (before a cut on this quantity). The
resolution on the invariant mass is 0.9 MeV/$c^2$. The tails of the
mass distribution originate from wrong photon pairing (the fraction
of these events estimated by a MC simulation is 0.2\%) and
$\pi\to\mu\nu$ decays. The background is negligible for the applied
mass cut.

It can be seen from (\ref{uvdef}) that the kinematic variable $u$
can be computed using only the $\pi^0\pi^0$ invariant mass. Thus a
measurement of $u$ uses the information from the LKr only, not
involving the DCH data. This provides a certain charge symmetry of
the procedure, as the LKr is a ``charge blind'' subdetector, except
for effects of small differences between $\pi^+$ and $\pi^-$
interaction characteristics.

\begin{figure}[tb]
\begin{center}
\vspace{-7mm}
\begin{tabular}{cc}
\resizebox{.5\textwidth}{!}{\includegraphics{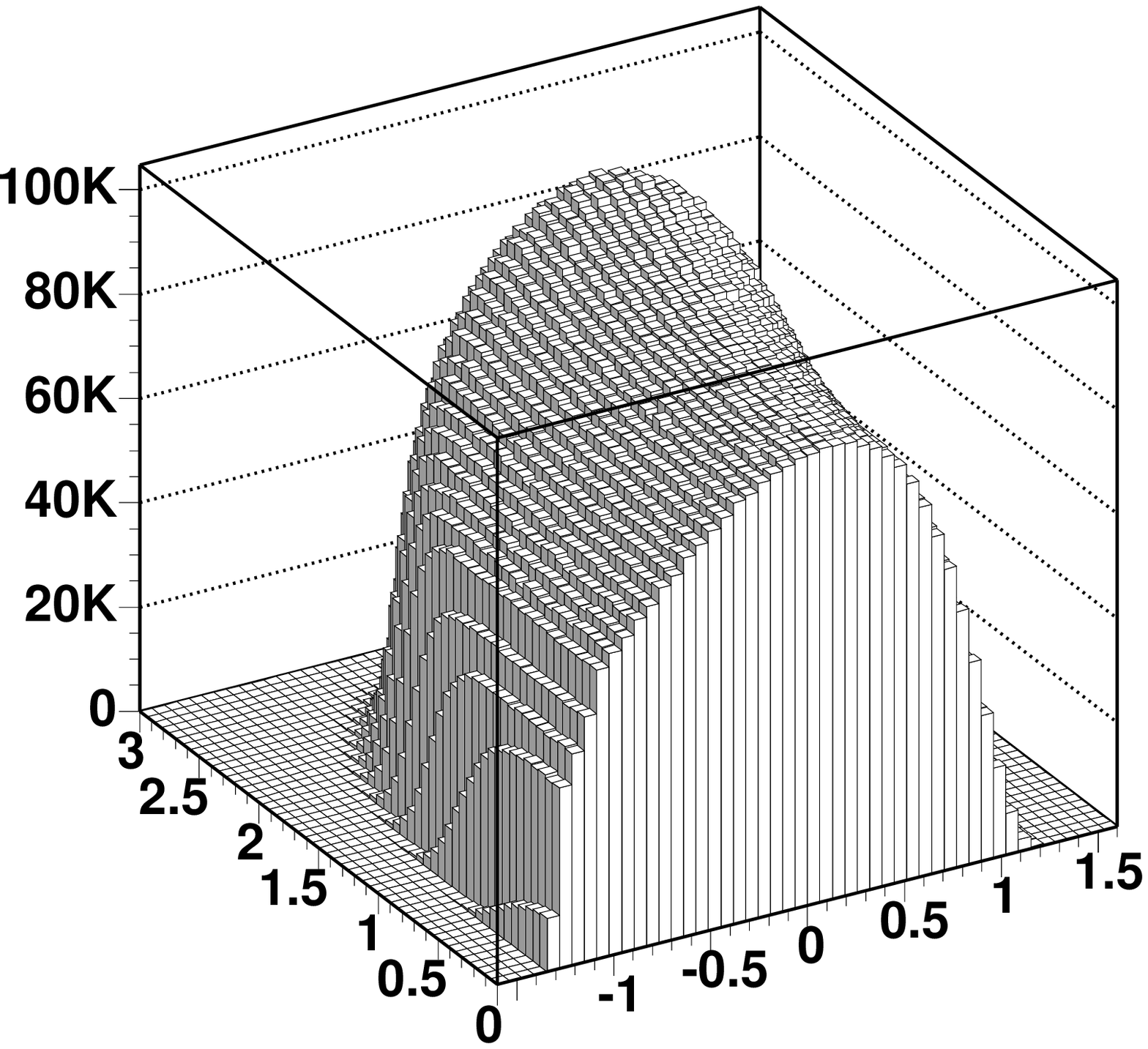}}
\put(-40,30){$u$} \put(-207,45){$|v|$} \hspace{0.5cm}
\resizebox{.45\textwidth}{!}{\includegraphics{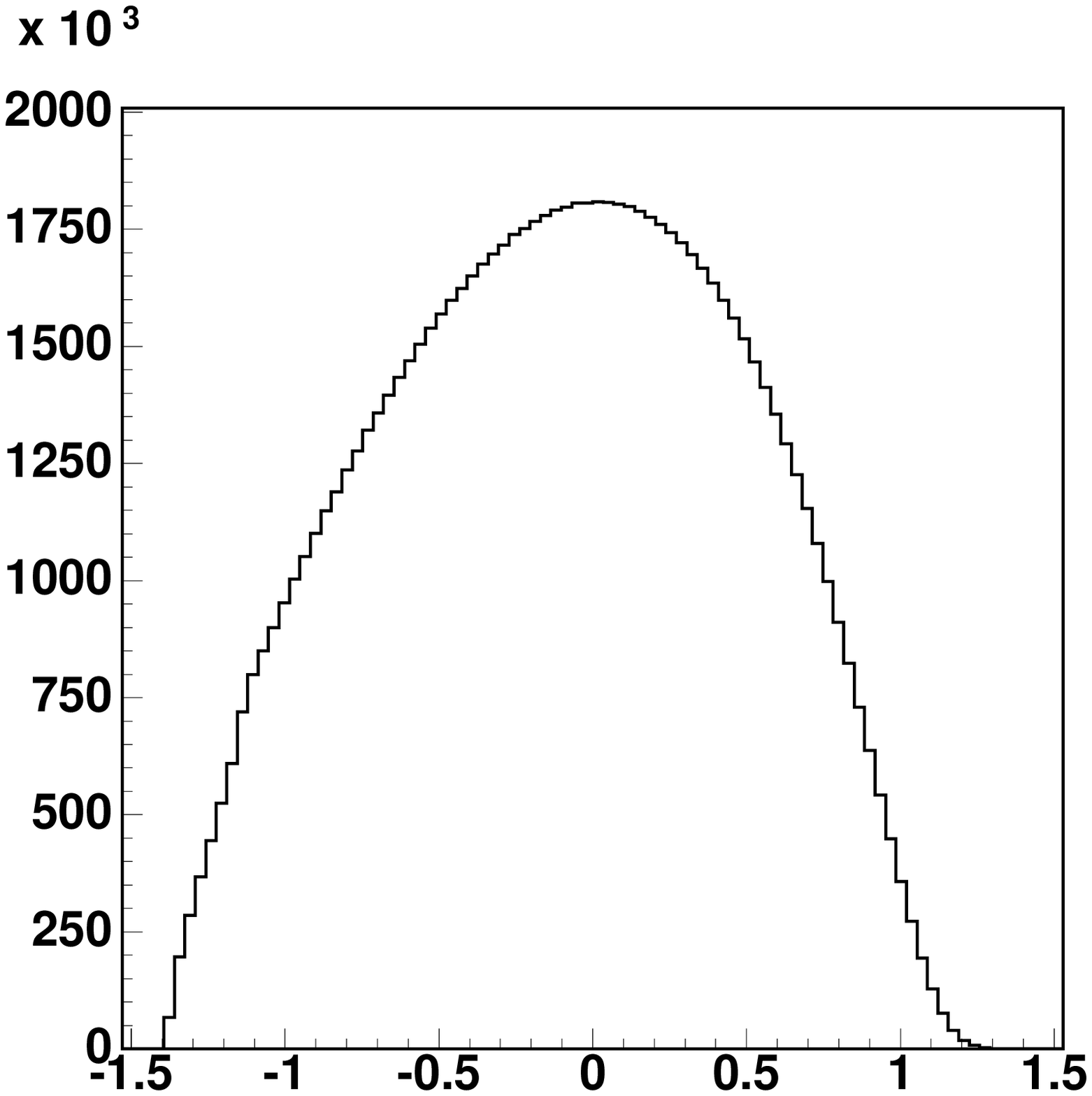}}
\put(-300,156){\Large\bf (a)} \put(-60,156){\Large\bf (b)}
\put(-25,5){$u$}\put(-215,140){\rotatebox{90}{Events}}
\end{tabular}
\end{center}
\vspace{-10mm} \caption{(a) Reconstructed Dalitz plot distribution
in the kinematic variables ($u$,$|v|$) for the selected
$K^\pm\to\pi^\pm\pi^0\pi^0$ events; (b) $u$-spectrum for the
selected events.} \label{dalitz}
\end{figure}

The reconstructed Dalitz plot distribution of the selected events is
shown in Fig.~\ref{dalitz}(a), and its projection on the $u$ axis is
presented in Fig.~\ref{dalitz}(b).

\boldmath
\subsection{Fits to $\Delta g^n$ and cross checks}
\unboldmath

Quadruple ratios of the $u$ spectra (\ref{quad}) were computed, and
$\Delta g^n$ was measured by the two methods described in
Section~\ref{sec:averaging} involving fitting with the function
(\ref{flin4}). The nominal values of slope parameters $g^n=0.626$,
$h^n=0.052$~\cite{pdg} were used. Unlike the
$K^\pm\to\pi^\pm\pi^+\pi^-$ case, no trigger corrections to the $u$
spectra are required, as discussed in Section~\ref{sec:neutrig}.

\begin{figure}[tb]
\begin{center}
{\resizebox*{0.7\textwidth}{!}{\includegraphics{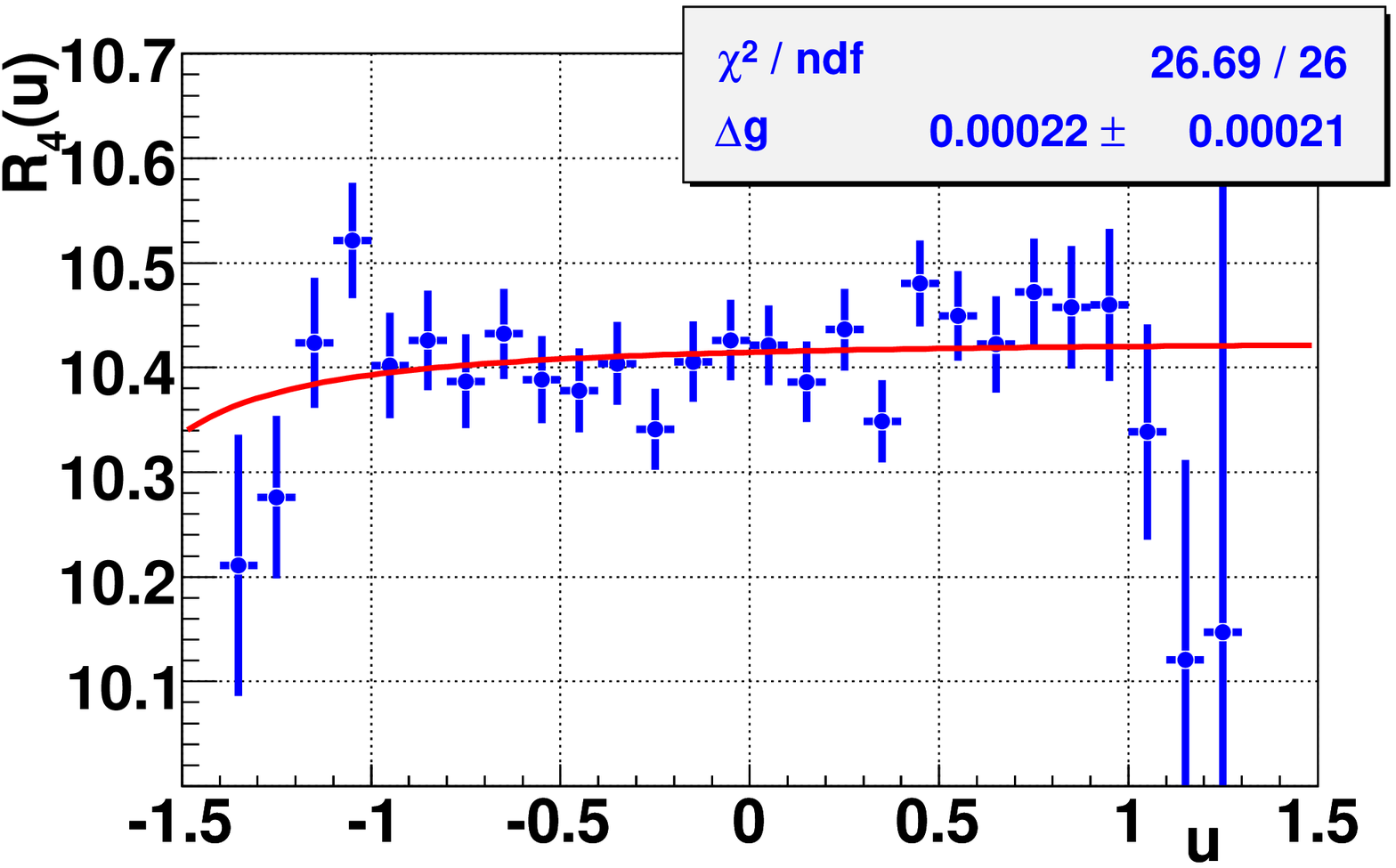}}}
\end{center}
\vspace{-9mm} \caption{The quadruple ratio $R^n_4(u)$ averaged by
supersamples in bins of $u$ fitted with the function (\ref{flin4})
with $g^n=0.626$, $h^n=0.052$~\cite{pdg}.} \label{fig:fitneu}
\end{figure}

A grand quadruple ratio obtained by averaging quadruple ratios over
supersamples in every bin of $u$ is tabulated in
Table~\ref{tab:fitneu}, and presented along with the result of the
corresponding fit in Fig.~\ref{fig:fitneu}.

\begin{table}[tb]
\caption{The quadruple ratio $R^n_4(u)$ averaged over supersamples.}
\begin{center}
\begin{tabular}{rrr|rrr}
\hline $u$ bin centre & Content & Error & $u$ bin centre & Content & Error\\
\hline
$-1.35$~~~~ & 10.2110 & 0.1248 & $0.05$~~~~ & 10.4213 & 0.0381 \\
$-1.25$~~~~ & 10.2762 & 0.0775 & $0.15$~~~~ & 10.3864 & 0.0382 \\
$-1.15$~~~~ & 10.4234 & 0.0621 & $0.25$~~~~ & 10.4363 & 0.0389 \\
$-1.05$~~~~ & 10.5216 & 0.0552 & $0.35$~~~~ & 10.3485 & 0.0393 \\
$-0.95$~~~~ & 10.4020 & 0.0504 & $0.45$~~~~ & 10.4804 & 0.0411 \\
$-0.85$~~~~ & 10.4261 & 0.0474 & $0.55$~~~~ & 10.4494 & 0.0430 \\
$-0.75$~~~~ & 10.3868 & 0.0449 & $0.65$~~~~ & 10.4222 & 0.0458 \\
$-0.65$~~~~ & 10.4322 & 0.0432 & $0.75$~~~~ & 10.4725 & 0.0507 \\
$-0.55$~~~~ & 10.3885 & 0.0415 & $0.85$~~~~ & 10.4576 & 0.0584 \\
$-0.45$~~~~ & 10.3781 & 0.0403 & $0.95$~~~~ & 10.4599 & 0.0726 \\
$-0.35$~~~~ & 10.4039 & 0.0395 & $1.05$~~~~ & 10.3386 & 0.1028 \\
$-0.25$~~~~ & 10.3408 & 0.0386 & $1.15$~~~~ & 10.1208 & 0.1912 \\
$-0.15$~~~~ & 10.4057 & 0.0383 & $1.25$~~~~ & 10.1472 & 0.5851 \\
$-0.05$~~~~ & 10.4262 & 0.0382 & $1.35$~~~~ &  8.8335 & 2.7018 \\
\hline
\end{tabular}
\end{center}
\label{tab:fitneu}
\end{table}

\begin{figure}[tb]
\begin{center}
\resizebox{0.81\textwidth}{!}{\includegraphics{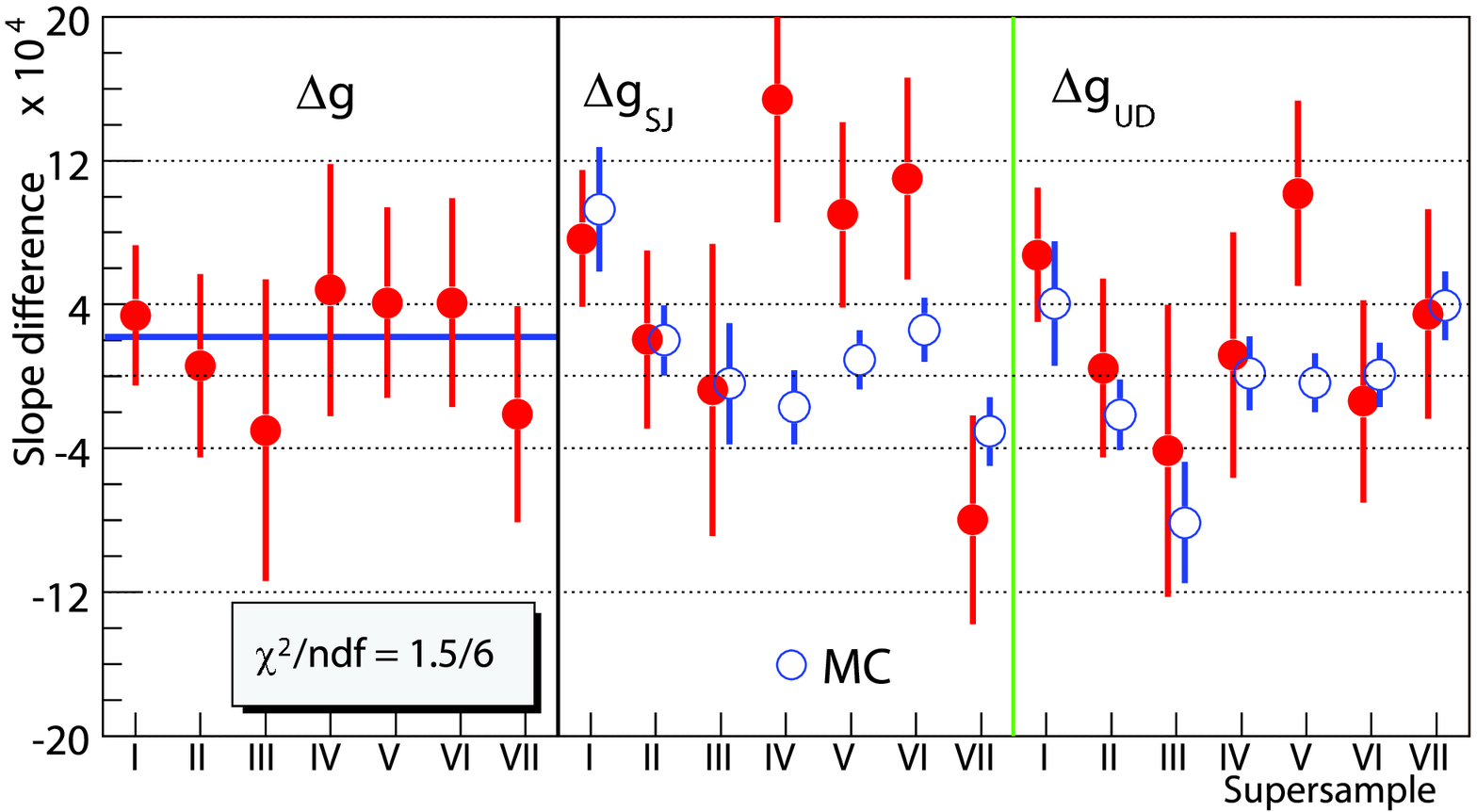}}
\put(-337,31){\Large\bf (a)} \put(-222,31){\Large\bf (b)}
\put(-108,31){\Large\bf (c)}
\end{center}
\vspace{-10mm} \caption{(a) $\Delta g^n$ measurement in the seven
supersamples; control
  quantities (b) $\Delta g_{SJ}^n$ and (c) $\Delta g_{UD}^n$ corresponding
  to detector and beam line asymmetries which cancel in quadruple
  ratio, and their comparison to MC.}
\label{fig:stabplotneu}
\end{figure}

Numbers of selected events and the results of the fits in every
supersample are presented in Table~\ref{tab:statsneu}. The
independent results obtained in the seven supersamples are shown in
Fig.~\ref{fig:stabplotneu}(a): the individual measurements are
compatible with a $\chi^2/{\rm ndf} = 1.5/6$.

\begin{table}[tb]
\caption{Selected statistics and measured $\Delta g^n$ in each
supersample.}
\begin{center}
\begin{tabular}{r|r|r|r}
\hline Supersample & $K^+\to\pi^+\pi^0\pi^0$&$K^-\to\pi^-\pi^0\pi^0$
&$\Delta g^n\times 10^4$\\
&decays in $10^6$&decays in $10^6$\\
\hline
I     & 16.40 & 9.14  & $3.4\pm 3.9$ \\
II    & 10.17 & 5.66  & $0.6\pm 5.1$ \\
III   &  3.71 & 2.06  & $-3.0\pm 8.4$ \\
IV    &  5.15 & 2.87  & $4.8\pm 7.1$ \\
V     &  8.88 & 4.94  & $4.1\pm 5.3$ \\
VI    &  7.49 & 4.17  & $4.1\pm 5.8$ \\
VII   &  6.86 & 3.82  & $-2.1\pm 6.0$ \\
\hline
Total & 58.66 & 32.66 & $2.2\pm 2.1$ \\
\hline
\end{tabular}
\end{center}
\label{tab:statsneu}
\end{table}

The measured control quantities $\Delta g_{SJ}^n$ and $\Delta
g_{UD}^n$, which are the slopes of the control ratios
(\ref{quad_LR}) and (\ref{quad_UD}), in the seven supersamples are
presented in Fig.~\ref{fig:stabplotneu}(b) and (c), respectively
(data points are overlayed with MC ones). These instrumental
asymmetries do not exceed the size of the statistical errors, and
are well reproduced by the MC simulation.

\newpage
\subsection{Systematic effects due to calorimeter performance}

As discussed above, the LKr calorimeter is the primary detector for
$K^\pm\to\pi^\pm\pi^0\pi^0$ decay reconstruction, and the only
detector used for measurement of the variable $u$. The only
charge-asymmetric effect in LKr performance is related to the
difference between $\pi^+$ and $\pi^-$ interaction cross sections.
Variation of the cut on the minimum allowed distance between the
photon candidate clusters and the impact point of the charged track
on LKr front face led to a conservative estimation of the
corresponding systematic uncertainty: $\delta(\Delta g^n)=0.5\times
10^{-4}$. Introduction of a shower energy sharing correction leads
to a consistent estimate.

Other effects related to the LKr reconstruction are not expected to
induce fake charge asymmetries. However the following stability
checks were performed.
\begin{itemize}
\item Stability with respect to variation of the resolution on the
reconstructed $u$ variable for $K^+$ and $K^-$ decays was studied
with a large sample of events simulated by MC. The result was found
to be stable within $0.1\times 10^{-4}$.
\item The result is stable to $0.1\times 10^{-4}$ with respect to
introducing various ways of correcting the measured photon energies
to account for nonlinearity of the LKr response at low energy.
\item The effect of wrong photon pairing in the reconstruction
of the $\pi^0\pi^0$ pair was studied by a large sample of simulated
events. The result was found stable to better than
$0.1\times10^{-4}$.
\item The result was found to have negligible sensitivy to variation
of the cluster radial distance cut value around the beam pipe.
\end{itemize}

\newpage
\subsection{Uncertainties due to trigger inefficiency}
\label{sec:neutrig}

The trigger logic is described in Section~\ref{sec:triggers}. Charge
symmetry of every trigger component was studied either by direct
measurement using samples recorded with low bias control triggers,
or by simulation, as discussed below.

The inefficiency of the HOD component of the L1N trigger, which is
due to inefficiency of the HOD counters, was measured using a
control sample of all events with exactly one reconstructed track,
triggered by conditions requiring activity in the LKr. The integral
inefficiency for the selected $K^\pm\to\pi^\pm\pi^0\pi^0$ sample was
measured to be about 0.25\%. It increased up to 2.0\% during short
periods due to malfunctioning of a few HOD counters; this effect was
reduced and symmetrized, as described in Section~\ref{sec:trigcor}.
The measured map of inefficiency measured as a function of the ($x$,
$y$) coordinates at the HOD plane allowed a precise estimation of
trigger inefficiency effect to be made. It was found to be
consistent with no spurious asymmetry at the level of $\delta(\Delta
g^n)=0.1\times 10^{-4}$.

The inefficiency of the LKr component of the L1N trigger was
measured with a sample of minimum bias triggers. It amounted to
0.7\% in supersamples I and II, and 3\% in supersamples III and IV.
For supersamples V, VI and VII the L1N condition was relaxed to
compensate for the above degradation by adding in ``OR'' a condition
requiring the total LKr energy deposition to exceed 15 GeV to the
initial condition based on the presence of at least two clusters,
which resulted in a stable inefficiency of 0.03\%. The degradation
of the trigger performance at the beginning of supersample III was
later identified to be due to a small time misalignment between
parts of the hardware trigger logic.

The inefficiency of the LKr component of L1N is a priori charge
symmetric, since it is based on LKr energy deposit conditions. To
confirm this, the inefficiency has been studied using a MC
simulation, with an LKr map of local trigger inefficiency (measured
directly from the data) used as a reference. Several checks have
been performed, in particular by artificially increasing the
measured inefficiency or by including totally inefficient regions.
No systematic effects have been observed at a level of
$\delta(\Delta g^n)=0.1\times10^{-4}$, which is considered a
systematic uncertainty due to the LKr component of the L1 trigger.

The inefficiency of the L2N trigger was mostly due to local
inefficiencies of the DCHs, and varied from 4$\%$ to 6$\%$. The
effects due to such inefficiencies, being of geometrical nature,
have been simulated by a MC. No charge asymmetry was found; upper
limits of systematic uncertainties from other (smaller) possible
effects, including those related to variations of timing offsets
between subdetectors and data buffer overflows, were estimated
directly by variation of the selection conditions. The total
systematic uncertainty induced by L2N inefficiency was estimated not
to exceed $\delta(\Delta g^n)=0.3\times10^{-4}$.

%\newpage
\subsection{Other systematic effects}
\label{sec:syst-neutral}

Effects related to the magnetic spectrometer do not affect the
result significantly, since the charged track is only used for the
identification of the kaon charge and in the mass cut. In
particular, systematic effects due to spectrometer misalignment (see
Section~\ref{sec:alignment}), momentum scale (see
Section~\ref{sec:spectrometer_field}), and geometrical acceptance
for the charged track (see Section~\ref{sec:beam-geom}), which are
important issues for the analysis in the $K^\pm\to3\pi^\pm$ mode,
were found to be negligible for the $K^\pm\to\pi^\pm\pi^0\pi^0$
analysis.

Uncertainties related to imperfect knowledge of the permanent
magnetic field in the decay volume were found not to exceed
$\delta(\Delta g^n)=0.1\times 10^{-4}$ by artificially varying the
field map in accordance with the precision of its measurement.
Stability of the result with respect to variation of the selected
$\pi^\pm\pi^0\pi^0$ invariant mass interval was checked, and no
systematic deviation was found.

Systematic uncertainty due to the difference of $K^+$ and $K^-$
production spectra was conservatively estimated in the same way as
for the $K^\pm\to\pi^\pm\pi^+\pi^-$ mode (described in
Section~\ref{sec:resid-charged}), and found to be $\delta(\Delta
g^n)=0.3\times 10^{-4}$.

Systematic uncertainties due to event pile-up were found not to
exceed $\delta(\Delta g^n)=0.2\times 10^{-4}$ by varying the
selection conditions on the allowed extra activity in the detectors,
and by checking the stability of the result with respect to the
timing cuts.

Charge-asymmetric material effects, which are negligible, were
discussed in Section~\ref{sec:resid-charged}.

\boldmath
\subsection{The resulting $A_g^n$}
\unboldmath

The systematic uncertainties are summarized in
Table~\ref{tab:systneu}. The difference in the linear slope
parameters of $K^+$ and $K^-$ decays into $\pi^\pm\pi^0\pi^0$ was
measured with the full NA48/2 data sample of $9.13\times 10^7$
events to be
\begin{equation}
\Delta g^n=(2.2\pm 2.1_{stat.}\pm 0.7_{syst.})\times 10^{-4}.
\label{result-neu}
\end{equation}
The corresponding direct CP violating asymmetry (\ref{agdef})
obtained using the nominal value of the linear slope parameter
$g^n=0.626\pm0.007$~\cite{pdg} is
\begin{equation}
A_g^n=(1.8\pm 1.7_{stat.}\pm 0.6_{syst.})\times 10^{-4}=
(1.8\pm1.8)\times 10^{-4}.
\end{equation}

\begin{table}[tb]
\caption{Systematic uncertainties on the measured value of $\Delta
g^n$.}
\begin{center}
\begin{tabular}{l|c}
\hline Systematic effect & Uncertainty $\delta(\Delta g^n)\times 10^4$\\
\hline
Overlap of LKr showers           & $\pm 0.5$ \\
L1 HOD trigger inefficiency      & $\pm 0.1$ \\
L1 LKr trigger inefficiency      & $\pm 0.1$ \\
L2 trigger inefficiency          & $\pm 0.3$ \\
Stray magnetic fields            & $\pm 0.1$ \\
Kaon production spectra          & $\pm 0.3$\\
Pile-up                          & $\pm 0.2$ \\
\hline
Total systematic uncertainty     & $\pm 0.7$ \\
\hline
\end{tabular}
\end{center}
\label{tab:systneu}
\end{table}

The uncertainties of the measured $A_g^c$ and $A_g^n$ are similar,
despite a ratio of the sample sizes of $N^c/N^n=34$. The main reason
for compensation of the difference in sample sizes is a small ratio
of the linear slope parameters of the two decay modes:
$|g^c/g^n|=0.34$.

%%%%%%%%%%%%%%%%%%%%%%%%%%%%%%%%%%%%%%%%%%%%%%%%%
%\newpage
\section{Discussion of the results}

The measurement of $\Delta g$ differences between the linear Dalitz
plot slopes was chosen as a representative quantity of a possible
direct CP violation effect, and is the quantity on which most
theoretical and experimental investigations are focused. In absence
of a specific model for physics beyond the SM it is not possible to
state in general terms which difference in $K^+$ and $K^-$ decay
distributions is expected to give the most significant CP violating
effect.

Discussing the asymmetry of linear slopes, it should be remarked
that the recent discovery by NA48/2~\cite{cusp} (and theoretical
interpretation~\cite{ca04,ca05,co06}) of the distortion of the
$\pi^\pm\pi^0\pi^0$ Dalitz plot distribution due to final state
interactions between pions indicated the need for alternative
parameterizations.
%The recent discovery by the NA48/2~\cite{cusp} and theoretical
%interpretation~\cite{ca04,ca05,co06} of the ``cusp effect'', which
%is a substantial modification of the $u$ spectrum of the
%$K^\pm\to\pi^\pm\pi^0\pi^0$ decays due to final state $\pi\pi$
%strong rescattering processes, led to development of a number of
%alternative parameterizations of the $K_{3\pi}^\pm$ decay
%distributions. One of them appears in the recent PDG
%review~\cite{pdg}.
Rescattering terms can be conveniently accommodated into a
parameterization based on a polynomial expansion of the decay
amplitude itself (rather than the event density) which, for the
non-rescattering part, would be:
\begin{equation}
|M'(u,v)|^2\sim (1+g'u/2+h'u^2/2+k'v^2/2+...)^2. \label{alt-slopes}
\end{equation}
The relation between the slope parameters appearing in
(\ref{slopes}) and (\ref{alt-slopes}) is, at first order, $g'=g$,
$h'=h-g^2/4$, $k'=k$. Measurement of $\Delta g'$ in the framework of
the parameterization (\ref{alt-slopes}) involves fitting of the
quadruple ratio of measured $u$ distributions (\ref{quad}) with a
function
\begin{equation}
f'(u)=n\cdot\left(1+\frac{ \frac{1}{2}\Delta
g'u}{1+\frac{1}{2}g'u+\frac{1}{2}h'u^2}\right)^8,
\end{equation}
rather than (\ref{flin4}).

The results of the measurement of $\Delta g$ in the framework of
parameterizations (\ref{slopes}) and (\ref{alt-slopes}) were
compared. The result in the $K^\pm\to3\pi^\pm$ mode is stable within
$0.1\times10^{-4}$, and its error is insensitive to the
parameterization. On the other hand, the result in the
$K^\pm\to\pi^\pm\pi^0\pi^0$ mode obtained with (\ref{alt-slopes}) is
\begin{equation}
\Delta g'=(3.2\pm3.1_{stat.})\times10^{-4}, \label{result-neu-alt}
\end{equation}
to be compared with (\ref{result-neu}). The significant variation of
the central value and its error is mostly due to the relatively
large values of the slope parameters $g^n$ and $h^n$. Introduction
of the additional $\pi\pi$ rescattering term to the amplitude
(\ref{alt-slopes}) does not considerably change the result with
respect to (\ref{result-neu-alt}).

The sensitivity discussed above is one of the reasons to publish the
tabulated quadruple ratios along with the results of the fits.

%%%%%%%%%%%%%%%%%%%%%%%%%%%%%%%%%%%%%%%%%%%%%%%%%%%%%%%%%%%
\section*{Conclusions}

NA48/2 has measured the charge asymmetries of Dalitz plot linear
slopes in both three-pion $K^\pm$ decay modes to be
\begin{equation}
A_g^c=(-1.5\pm2.2)\times 10^{-4},~~~A_g^n=(1.8\pm1.8)\times 10^{-4},
\end{equation}
which is an improvement in accuracy over the previous
measurements~\cite{pdg} by more than one order of magnitude. NA48/2
precisions are limited mainly by the available statistics.

The measured asymmetries do not show evidences for large
enhancements due to non-SM physics. They are consistent with the SM
predictions, in particular with a full next-to-leading order ChPT
calculation~\cite{ma95} predicting
\begin{equation}
A_g^c=(-1.4\pm1.2)\times 10^{-5},~~~A_g^n=(1.1\pm0.7)\times 10^{-5}.
\end{equation}
Due to the high precision achieved, the results can be used to
constrain extensions of the SM predicting enhancements of the CP
violating effects.

%%%%%%%%%%%%%%%%%%%%%%%%%%%%%%%%%%%%%%%%%%%%%%%%%%%%%%%%%%%
\section*{Acknowledgements}

It is a pleasure to thank the technical staff of the participating
laboratories, universities and affiliated computing centres for
their efforts in the construction of the NA48 apparatus, in the
operation of the experiment, and in the data processing.

%\end{linenumbers}

%

%

\begin{thebibliography}{99}
%
\bibitem{ch64}
J.H. Christenson {\it et al}., Phys. Rev. Lett. {\bf 13} (1964) 138.
%
\bibitem{ba93}
H. Burkhardt {\it et al}. (NA31), Phys. Lett. {\bf B206} (1988) 169.\\
G. Barr {\it et al}. (NA31), Phys. Lett. {\bf B317} (1993) 233.
%
\bibitem{fa99}
V. Fanti {\it et al}. (NA48), Phys. Lett. {\bf B465} (1999) 335.\\
A. Lai {\it et al}. (NA48), Eur. Phys. J. {\bf C22} (2001) 231.\\
J.R. Batley {\it et al}. (NA48), Phys. Lett. {\bf B544} (2002) 97.
%
\bibitem{al99}
A. Alavi-Harati {\it et al}. (KTeV), Phys. Rev. Lett. {\bf 83} (1999) 22.\\
A. Alavi-Harati {\it et al}. (KTeV), Phys. Rev. {\bf D67} (2003)
012005.
%
\bibitem{au01}
B. Aubert {\it et al}. (Babar), Phys. Rev. Lett. {\bf 87} (2001) 091801.\\
K. Abe {\it et al}. (Belle), Phys. Rev. Lett. {\bf 87} (2001)
091802.
%
\bibitem{ab04}
K. Abe {\it et al}. (Belle), Phys. Rev. Lett. {\bf 93} (2004) 021601.\\
B. Aubert {\it et al}. (Babar), Phys. Rev. Lett. {\bf 93} (2004)
131801.
%
\bibitem{pdg}
W.-M. Yao {\it et al}. (PDG), J. Phys. {\bf G33} (2006) 1.
%
\bibitem{ca04}
N. Cabibbo, Phys. Rev. Lett. {\bf 93} (2004) 121801.
%
\bibitem{ca05}
N. Cabibbo and G. Isidori, JHEP {\bf 0503} (2005) 021.
%
\bibitem{co06}
G. Colangelo {\it et al.}, Phys. Lett. {\bf B638} (2006) 187.
%
\bibitem{is92}
G. Isidori, L. Maiani, A. Pugliese, Nucl. Phys. {\bf B381} (1992)
522.
%
\bibitem{ma95}
E. G\'amiz, J. Prades, I. Scimemi, JHEP {\bf 10} (2003) 042.
%
\bibitem{fa05}
G. F\"aldt, E. Shabalin, Phys. Lett. {\bf B635} (2006) 295.
%
\bibitem{sh98}
E.P. Shabalin, ITEP preprint {\bf 8-98} (1998).\\
G. D'Ambrosio, G. Isidori, G. Martinelli, Phys. Lett. {\bf B480}
(2000) 164.
%
\bibitem{fo70}
W.T. Ford {\it et al}., Phys. Rev. Lett. {\bf 25} (1970) 1370.
%
\bibitem{sm75}
K.M. Smith {\it et al}., Nucl. Phys. {\bf B91} (1975) 45.\\
G.A. Akopdzhanov {\it et al}. (TNF--IHEP), Eur. Phys. J. {\bf C40}
(2005) 343.
%
\bibitem{k3pi}
J.R. Batley {\it et al.} (NA48/2), Phys. Lett. {\bf B634} (2006)
474.
%
\bibitem{k3pi-n}
J.R. Batley {\it et al.} (NA48/2), Phys. Lett. {\bf B638} (2006) 22.
%
\bibitem{gi96}
Y. Giomataris {\it et al.}, \NIMA{376}{1996}{29}.
%
\bibitem{pe04}
B. Peyaud {\it et al.}, \NIMA{535}{2004}{247}.
%
\bibitem{bed95}
D. B\`eder\'ede {\it et al.}, \NIMA{367}{1995}{88}.
%
\bibitem{ba96}
G.D. Barr {\it et al.}, \NIMA{370}{1996}{413}.
%
\bibitem{fa07}
V. Fanti {\it et al.} (NA48), \NIMA{574}{2007}{433}.
%
\bibitem{slopes}
J.R. Batley {\it et al.} (NA48/2), Phys. Lett. {\bf B649} (2007)
349.
%
\bibitem{geant}
GEANT Description and Simulation Tool, CERN Program Library Long
Writeup {\bf W5013} (1994).
%
\bibitem{fr87}
R. Fr\"uhwirth, \NIMA{262}{1987}{444}.
%
\bibitem{cu02}
J.R. Cudell {\it et al.}, Phys. Rev. {\bf D65} (2002) 074024.
%
\bibitem{at80}
H.W. Atherton {\it et al.}, CERN 80-07.
%
\bibitem{cusp}
J.R. Batley {\it et al.} (NA48/2), Phys. Lett. {\bf B633} (2006)
173.
%
\end{thebibliography}
\end{document}